\documentclass[11pt]{article}

\title{Classification of Solutions in Topologically Massive Gravity}

\usepackage{amsmath}
\usepackage{amsfonts}
\usepackage{amssymb}
\usepackage{amscd}
\usepackage{latexsym}
\usepackage{setspace}
\usepackage{rotating}

\spacing{1.4}

\makeatletter
\@addtoreset{equation}{section}
\makeatother

\newcommand{\w}[1]{\\[0.#1cm]}
\def\eq#1{(\ref{#1})}

\newcommand{\be}{\begin{equation}}
\newcommand{\ee}{\end{equation}}
\newcommand{\ben}{\begin{equation}}
\newcommand{\een}{\end{equation}}
\newcommand{\bea}{\setlength\arraycolsep{2pt} \begin{eqnarray}}
\newcommand{\eea}{\end{eqnarray}}
\newcommand{\nn}{\nonumber}
\newcommand{\nnr}{\nonumber \\}
\newcommand{\la}{\label}

\newcommand{\bib}{\bibitem}
\newcommand{\ci}{\cite}
\newcommand{\se}{\section}
\newcommand{\sse}{\subsection}
\newcommand{\ssse}{\subsubsection}
\newcommand{\pa}{\paragraph}
\newcommand{\qd}{\quad}

\newcommand{\lt}{\left}
\newcommand{\rt}{\right}
\newcommand{\fr}{\frac}

\newcommand{\tf}{\tfrac}
\newcommand{\ft}[2]{{\textstyle{\frac{\scriptstyle #1}{\scriptstyle #2}}}}

\newcommand{\leri}{\leftrightarrow}

\newcommand{\na}{\nabla}

\newcommand{\pd}{\partial}
\newcommand{\ra}{\rightarrow}

\newcommand{\sr}{\sqrt}
\newcommand{\we}{\wedge}

\newcommand{\ga}{\alpha}
\newcommand{\gb}{\beta}
\newcommand{\gc}{\gamma}

\newcommand{\gd}{\delta}

\newcommand{\gep}{\epsilon}
\newcommand{\ep}{\epsilon}
\newcommand{\gve}{\varepsilon}

\newcommand{\gy}{\eta}

\newcommand{\gq}{\theta}

\newcommand{\gk}{\kappa}
\newcommand{\gl}{\lambda}
\newcommand{\gL}{\Lambda}

\newcommand{\gr}{\rho}

\newcommand{\gs}{\sigma}

\newcommand{\gt}{\tau}

\newcommand{\gf}{\phi}

\newcommand{\gvf}{\varphi}

\newcommand{\gw}{\omega}

\newcommand{\ud}{\textrm{d}}
\newcommand{\ue}{\textrm{e}}

\newcommand{\ui}{\textrm{i}}

\newcommand{\us}{\textrm{s}}
\newcommand{\ut}{\textrm{t}}

\newcommand{\im}{\textrm{i}}

\newcommand{\uSO}{\textrm{SO}}

\newcommand{\uSU}{\textrm{SU}}

\newcommand{\tr}{\textrm{tr}}

\newcommand{\mI}{\mathsf{I}}
\newcommand{\mJ}{\mathsf{J}}

\newcommand{\mM}{\mathsf{M}}

\newcommand{\mS}{\mathsf{S}}

\newcommand{\bbC}{\mathbb{C}}

\newcommand{\bbH}{\mathbb{H}}

\newcommand{\bbR}{\mathbb{R}}

\newcommand{\bbZ}{\mathbb{Z}}

\newcommand{\cA}{\mathcal{A}}

\newcommand{\cF}{\mathcal{F}}

\newcommand{\ds}{\textrm{d} s^2}

\newcommand{\del}{\partial}

\begin{document}

\begin{titlepage}
\begin{flushright}
MIFP-09-27
\end{flushright}
\vspace*{100pt}
\begin{center}
{\bf \Large{Classification of Solutions in Topologically Massive Gravity}}\\
\vspace{50pt}
\long\def\symbolfootnote[#1]#2{\begingroup
\def\thefootnote{\fnsymbol{footnote}}\footnote[#1]{#2}\endgroup}
\large{David D.K. Chow$^1$, C.N. Pope$^{1, 2}$ and Ergin Sezgin$^1$}
\end{center}

\begin{center}

$^1${\it George P. and Cynthia W. Mitchell Institute for
Fundamental Physics \& Astronomy,\\
Texas A\&M University, College Station, TX 77843-4242, USA}\\
\vspace*{10pt}
$^2${\it Department of Applied Mathematics and Theoretical Physics, University
of Cambridge,\\
Centre for Mathematical Sciences, Wilberforce Road, Cambridge CB3 0WA, UK}\\

\vspace{50pt}

{\bf Abstract}

\end{center}

\noindent We study exact solutions of three-dimensional gravity with a cosmological constant and a gravitational Chern--Simons term: the theory known as topologically massive gravity.  After reviewing the algebraic classification, we show that if a solution has curvature of algebraic type D, then it is biaxially squashed AdS$_3$.  Applying the classification, we provide a comprehensive review of the literature, showing that most known solutions are locally equivalent to biaxially squashed AdS$_3$ or to AdS pp-waves.

\end{titlepage}

\tableofcontents

\newpage


\se{Introduction}


Three-dimensional gravity with a gravitational Chern--Simons term, and possibly
with a cosmological constant, is known as topologically massive gravity (TMG)
\cite{Deser:1982vya, Deser:1982vyb, Deser:1982sv},
and is described by the action
\be
I= \frac1{16\pi G} \int \ud^3 x \, \sqrt{-g} \left[ R-2\Lambda
 + \frac{1}{2\mu} \gep^{\lambda\mu\nu}
\Gamma{^\rho}_{\lambda\sigma} \left( \partial_\mu\Gamma{^\sigma}_{\rho\nu}
+\frac23 \Gamma{^\sigma}_{\mu\tau} \Gamma{^\tau}_{\nu\rho}\right) \right]\, ,
\label{1.1}
\ee
where $G$ is Newton's constant, $\Lambda$ is the cosmological constant and $\mu$
is a mass parameter.
The model supports a black hole solution \cite{Banados:1992wn, baheteza} for
$\Lambda<0$, and for a generic value of the Chern--Simons coupling constant
$\mu$ it has a propagating massive graviton.  The theory also admits a dual
boundary conformal field theory description.  All these properties make the
model a natural one in which to study various aspects of non-perturbative
gravity that seem to be forbiddingly difficult in higher dimensions.

In Einstein gravity, for which the Chern--Simons term is absent, and with a
negative cosmological constant, there have been attempts to solve the theory
exactly \cite{Witten:1988hc, Witten:2007kt, Maloney:2007ud}, by using knowledge
of the solution space for fixed boundary conditions. Certain problems
encountered in giving a physical interpretation to the partition function
\cite{Maloney:2007ud} might be circumvented by considering TMG instead, and, in
particular, at a critical value of the Chern--Simons coupling
constant: $\mu=\sqrt{ -\Lambda}$ \cite{LSS}.  At this chiral point, with the
further assumptions that $G>0$, and that the standard Brown--Henneaux boundary
conditions \cite{Brown:1986nw} hold, the theory is known as chiral gravity. 
This proposal motivates an exhaustive investigation of the solution space of
TMG, in which the Chern--Simons coupling plays a non-trivial r\^{o}le.

The field equation for TMG, obtained by varying (\ref{1.1}) with respect to the
metric, is
\be
R_{\mu \nu} - \fr{1}{2} R g_{\mu \nu}
+ \gL g_{\mu \nu} + \fr{1}{\mu} C_{\mu \nu} = 0 \, ,
\la{fieldeq1}
\ee
where $C_{\mu \nu}$ is the Cotton tensor, which is a symmetric and
traceless tensor defined as
\ben
C_{\mu \nu} = \gep{_\mu}{^{\gr \gs}}
\na_\gr (R_{\gs \nu} - \tf{1}{4} R g_{\gs \nu}) \, .
\een
$\gep_{\mu \nu \gr}$
is the Levi-Civita tensor, related by $\gep_{\mu \nu \gr} = \sr{-g} \gve_{\mu
\nu \gr}$ to the weight $+1$ tensor density $\gve_{\mu \nu \gr}$.  Strictly
speaking, the Cotton and Levi-Civita tensors are pseudotensors --- they change
sign under a parity transformation --- and so the field equation is not quite
tensorial.  When specifying any solution, it is therefore essential to state
both the metric and the convention for $\gep_{\mu \nu \gr}$; we take $\gve_{012}
= + 1$.  We usually consider the cosmological constant to be 
negative: $\gL = - m^2$, where $|m|$ is the inverse AdS radius.  However, we
shall sometimes consider a vanishing cosmological constant or, by taking $m \ra
\ui m$, a positive cosmological constant $\gL = m^2$.  From the trace of the
field equation, we deduce that $R = 6 \gL$, and so the field equation can be
written more simply as
\ben
R_{\mu \nu} - 2 \gL g_{\mu \nu} +
\fr{1}{\mu} \gep{_\mu}{^{\gr \gs}} \na_\gr R_{\gs \nu} =0 \, . \la{fieldeq2}
\een
Henceforth, it suffices to consider only this form of the field equation.

We shall only consider local solutions of TMG.  In three dimensions, the Riemann and Ricci tensors both have 6 independent components; these tensors are related by
\ben
R_{\mu \nu \gr \gs} = 2 R_{\mu [ \gr} g_{\gs ] \nu} -
2 R_{\nu [ \gr} g_{\gs ] \mu} - R g_{\mu [\gr} g_{\gs ] \nu}\, .
\la{Riem3d}
\een
Any Einstein metric is therefore maximally symmetric, i.e.~locally flat for $\Lambda = 0$, anti-de Sitter for $\Lambda < 0$, or de Sitter for $\Lambda>0$.  These are automatically solutions of TMG, since they have a vanishing Cotton tensor; in fact, a solution of TMG has a vanishing Cotton tensor if and only if it is Einstein with $R_{\mu \nu} = 2 \gL g_{\mu \nu}$.  

We henceforth consider non-trivial exact solutions of TMG that have a non-vanishing Cotton tensor, which are not Einstein, but it turns out that remarkably few are known.  Although there is much
literature about exact solutions of TMG, almost all local metrics reduce to
three particular solutions:
\begin{enumerate}

\item Timelike-squashed AdS$_3$ \cite{nutku, Gurses:1994}, which is a biaxially
squashed AdS$_3$ and can be regarded as a timelike fibration over $\mathbb{H}^2$.

\item Spacelike-squashed AdS$_3$ \cite{boucle}, which is a biaxially squashed
AdS$_3$ and can be regarded as a spacelike fibration over AdS$_2$.

\item AdS pp-waves \cite{ayohas1, olsate, ayohas3}, which are generalizations of
pp-waves to include a cosmological constant.

\end{enumerate}
The only solutions in the literature that are not locally
(special cases of) timelike- or spacelike-squashed AdS$_3$ or an AdS pp-wave are the general triaxially squashed AdS$_3$ solutions of Nutku and Baekler
\ci{nutbak}, given in their equations (4.1), (4.6) and (4.8), also found by
Ortiz \ci{ortiz}, given in his equation (5.4).  These exceptional triaxially
squashed solutions solve TMG without a cosmological constant, generalizing the
timelike- and spacelike-squashed AdS$_3$ solutions, which have only biaxial
squashing.

In this paper, we only consider local solutions, and do not distinguish between different global identifications.  The global solutions of the theory without the Chern--Simons term are well-known; they
are obtainable by making conformal transformations of the AdS$_3$ metric that
preserve the Brown--Henneaux asymptotics (see, for example,
\cite{Banados:1998gg} for a review).  These include the BTZ black hole \cite{Banados:1992wn, baheteza}, which is obtained from discrete quotienting of AdS$_3$.  Similarly, discrete quotients of the squashed AdS$_3$ solutions of TMG give black holes \cite{nutku, clement, Gurses:1994, aiclle, boucle}.  Recently, these black holes were systematically studied in \cite{anlipasost}\footnote{Because the squashing parameters are constants, we call them squashed AdS$_3$ black holes, rather than ``warped'' AdS$_3$ black holes as referred to by \cite{anlipasost}.}.

As in four-dimensional general relativity, one of the problems when surveying
the literature of exact solutions is to disentangle genuinely new solutions from those that are already known but written in different coordinate systems.  One
approach to bringing some order to the compilation is to use the algebraic
classification of curvature.  We can classify the Cotton
tensor \cite{babula, Hall:1999, Torres:2003} (see also \cite{gahehema,
Sousa:2007ax}), analogous to the four-dimensional Petrov classification of the Weyl tensor.  Alternatively, we can classify the
three-dimensional traceless Ricci tensor \cite{hamope, Hall:1999, Torres:2003}
(see also \cite{Sousa:2007ax}), analogous to the Segre classification of the energy-momentum tensor in four-dimensional general relativity.  In fact, a consequence of the field equation is
that the traceless Ricci tensor $S_{\mu \nu} = R_{\mu \nu} - \tf{1}{3} R g_{\mu
\nu}$ and the Cotton tensor are proportional:
\ben
S_{\mu \nu} + \fr{1}{\mu} C_{\mu \nu} = 0 \, . \la{SCprop}
\een
Thus the Petrov classification of the Cotton tensor and the Segre classification
of the traceless Ricci tensor are equivalent in TMG; we shall refer to it as the
``Petrov--Segre classification''.  In this paper, we survey all known solutions of TMG, determining their Petrov--Segre types and equivalences under local coordinate transformations, and study to what degree an exact solution can be determined by specifying its Killing symmetries and its Petrov--Segre type.

The problem of finding all solutions of TMG with a given set of
Killing symmetries and with a specified Petrov--Segre class is in general
highly non-trivial.  A noteworthy exception is that by requiring the existence
of a null Killing vector $k$, the solutions to TMG can be shown to be of
Petrov--Segre type N, with the traceless Ricci tensor $S_{\mu\nu}$ proportional
to $k_\mu\, k_\nu$ \cite{gipose}.  These solutions are known as AdS pp-waves
\cite{ayohas1}, and they have been rediscovered on a number of occasions.

Here we shall also show that a Petrov--Segre type D solution of TMG must be a
biaxially squashed AdS$_3$.  Our main application
of this result is that it provides an algorithm for testing whether a solution
is a biaxially squashed AdS$_3$ in some coordinate
system.  The power of this result is illustrated in Appendices \ref{A1} and
\ref{A2}, where we show that various solutions in the literature are, by
coordinate transformation, biaxially squashed AdS$_3$ in disguise, despite a
sm\"{o}rg\r{a}sbord of coordinate systems to contend with.

The outline of the rest of this paper is as follows.  In Section 2, we review the Petrov--Segre algebraic
classification of curvature in three dimensions.  In Section 3, we prove that type D spacetimes that are solutions of TMG must be squashed AdS$_3$.  We complete the enumeration of the known solutions of TMG with a review of the pp-wave solution in Section 4.  After presenting our conclusions in Section 5, we give, in the Appendix, a comprehensive review of the literature of previously known TMG
solutions, showing in particular that most are related by local coordinate
transformations to the squashed AdS$_3$ and AdS pp-wave solutions.


\se{Algebraic classification of curvature}


A considerable effort has gone into finding exact solutions of
four-dimensional general relativity \ci{stkrmahohe}.  There are four
main classification schemes: symmetry groups, such as isometry groups; algebraic
classification of the Weyl tensor (Petrov classification); algebraic
classification of the traceless Ricci tensor (Segre classification); and
solutions admitting special vector or tensor fields that satisfy certain
geometrically meaningful differential constraints.  These ideas motivate
approaches to finding and classifying exact solutions of TMG.  A difference, specific to TMG, is that the three-dimensional analogues of algebraically
classifying curvature, which are to classify the Cotton tensor and the traceless
Ricci tensor, coincide in view of \eq{SCprop}.  In this paper, we focus on algebraic classification and, to a lesser extent, symmetry; in a subsequent paper \cite{chposeII}, the focus is on solutions with special null vector fields.


\sse{Comparison of three and four dimensions}


In four-dimensional general relativity, the 20 independent components
of the Riemann tensor split into: 10 independent components of the
Weyl tensor $C_{\mu \nu \gr \gs}$; 9 independent components of the
traceless Ricci tensor $S_{\mu \nu} = R_{\mu \nu} - \tf{1}{4} R g_{\mu
  \nu}$; and 1 component of the Ricci scalar $R$.  There are two
independent algebraic classifications of curvature: the Petrov
classification of the Weyl tensor (see, for example, Chapter 4 of
\ci{stkrmahohe}), and the Segre (or Pleba\'{n}ski) classification of
the traceless Ricci tensor (see, for example, Chapter 5 of
\ci{stkrmahohe}).  There are several formulations, but one is that they classify according to the eigenvalues of
linear maps associated with the curvature tensors, in particular the
algebraic and geometric multiplicities of the eigenvalues.

In four dimensions, the 10 real components of the Weyl tensor can be packaged
into a traceless and symmetric $3 \times 3$ complex matrix $C_{ab}$.  One
formulation of the four-dimensional Petrov classification is algebraic classification of
$C{^a}{_b}$, regarded as a linear map between complex 3-vectors.  However, in three
dimensions the Riemann tensor can be expressed in terms of the Ricci tensor, so the Weyl tensor vanishes and there is no direct analogue of the
Petrov classification.  Instead, the r\^{o}le of the conformal tensor in three
dimensions is played by the Cotton tensor; it is analogous to the Weyl tensor in
four dimensions and higher, in that its vanishing is equivalent to conformal
flatness.  A three-dimensional analogue of the Petrov classification is
algebraic classification of the Cotton tensor $C{^a}{_b}$, regarded as a linear
map between 3-vectors.  The three-dimensional Petrov classification of the
Cotton tensor has been studied in \ci{babula, Hall:1999, Torres:2003, gahehema,
Sousa:2007ax}.

In four dimensions, the Segre classification is algebraic classification of the
traceless Ricci tensor $S{^a}{_b} = R{^a}{_b} - \tf{1}{4} R \gd^a_b$, regarded
as a linear map between 4-vectors.  By the Einstein equation, this is equivalent
to algebraic classification of the energy-momentum tensor.  There is a natural
three-dimensional analogue: algebraic classification of $S{^a}{_b} = R{^a}{_b} -
\tf{1}{3} R \gd^a_b$, regarded as a linear map between 3-vectors.  The
three-dimensional Segre classification of the traceless Ricci tensor has been
studied in \ci{hamope, Hall:1999, Torres:2003, Sousa:2007ax}.


\sse{Petrov--Segre classification in TMG}


For a general three-dimensional theory, these analogues of the Petrov and Segre
classifications are distinct.  However, in TMG the Cotton tensor and the
traceless Ricci tensor are proportional (\ref{SCprop}), so the two
classifications coincide; in this context, we call it the Petrov--Segre
classification.  It might be better to regard it as classification of the traceless Ricci tensor, since the Cotton tensor comes from differentiation of the Ricci tensor and so is less fundamental, however the classification is algebraic.

\begin{sidewaystable}[htbp]
\begin{center}
\begin{tabular}{c|c|c|c|c|c}
Petrov & Segre & Canonical $S_{ab}$ & $S{^a}{_b}$
Jordan form & Jordan basis & Minimal polynomial \\
\hline
O & $[(11,1)]$ & $0$ & $0$ & $(l^b, n^b, m^b)$ & $\mS$ \\
\hline
N & $[(12)]$ & $\lambda l_a l_b$ &
$\begin{pmatrix}
0 & 1 & 0 \\
0 & 0 & 0 \\
0 & 0 & 0
\end{pmatrix}$
& $(l^b, - \gl n^b, m^b)$ & $\mS^2$ \\
\hline
D$_\ut$ & $[(11),1]$ & $\alpha (\eta_{ab} + 3t_a t_b )$ &
$\begin{pmatrix}
\ga & 0 & 0 \\
0 & \ga & 0 \\
0 & 0 & - 2 \ga
\end{pmatrix}$
& $(m^b, z^b, t^b)$ & $(\mS + 2 \ga \mI) (\mS - \ga \mI)$ \\
\hline
D$_\us$ & $[1(1,1)]$ & $\alpha (\eta_{ab} -3 m_a m_b )$ &
$\begin{pmatrix}
\ga & 0 & 0 \\
0 & \ga & 0 \\
0 & 0 & - 2 \ga
\end{pmatrix}$
& $(t^b, z^b, m^b)$ & $(\mS + 2 \ga \mI) (\mS - \ga \mI)$ \\
\hline
III & $[3]$ & $2 \tau  l_{(a} m_{b)}$ &
$\begin{pmatrix}
0 & 1 & 0 \\
0 & 0 & 1 \\
0 & 0 & 0
\end{pmatrix}$
& $(l^b, \gt m^b, -n^b)$ & $\mS^3$ \\
\hline
II & $[12]$ &
$\begin{array}{c} \alpha (\eta_{ab} -3 m_am_b) \\ + \lambda l_a l_b \end{array}$
&
$\begin{pmatrix}
\ga & 1 & 0 \\
0 & \ga & 0 \\
0 & 0 & - 2 \ga
\end{pmatrix}$
& $(l^b, - \gl n^b, m^b)$ & $(\mS + 2 \ga \mI) (\mS - \ga \mI)^2$ \\
\hline
I$_\bbR$ & $[11,1]$ & $\begin{array}{c} \alpha (\eta_{ab} -3 m_am_b) \\ -\beta\,
(l_a l_b + n_a n_b ) \end{array}$ & 
$\begin{pmatrix}
\ga + \gb & 0 & 0 \\
0 & \ga - \gb & 0 \\
0 & 0 & - 2 \ga
\end{pmatrix}$
& $(l^b+n^b, l^b-n^b, m^b)$ & $(\mS + 2 \ga \mI) (\mS - \ga \mI + \gb \mI) (\mS
- \ga \mI - \gb \mI)$ \\
\hline
I$_\bbC$ & $[1 z {\bar z}]$ & $\begin{array}{c} \alpha (\eta_{ab} -3 m_am_b) \\
-\beta\, (l_a l_b -n_a n_b ) \end{array}$ &
$\begin{pmatrix}
\ga + \im \gb & 0 & 0 \\
0 & \ga - \im \gb & 0 \\
0 & 0 & - 2 \ga
\end{pmatrix}$
& $(l^b + \im n^b, l^b - \im n^b, m^b)$ & $(\mS + 2 \ga \mI) (\mS - \ga \mI +
\im \gb \mI) (\mS - \ga \mI - \im \gb \mI)$ \\
\end{tabular}
\end{center}
\caption{Petrov--Segre classification.  This shows, for each Petrov--Segre type,
the canonical form of the traceless Ricci tensor, its Jordan normal form, and
the monic polynomial of $S{^a}{_b}$ of minimal degree that vanishes.  Here,
$\ga$ and $\gb$ are functions of spacetime, $\gl$ and $\gt$ are constants; $l$
and $n$ are null, $m$ and $z$ are spacelike, $t$ is timelike; the matrix $\mS$
has entries $S{^a}{_b}$.
}
\label{table1}
\end{sidewaystable}

The Petrov--Segre classification is given in Table \ref{table1}.  $(l^a, m^a,
n^a)$ represents a null frame: $l$ and $n$ are null vectors, $m$ is a unit
spacelike vector, and their only non-vanishing inner products are $l^a n_a=-1$
and $m^a m_a=1$, so $\gy_{ab} = - l_a n_b - n_a l_b + m_a m_b$ is flat.  We can also
define the unit timelike vector $t^a=(l^a+n^a)/{\sqrt 2}$ and unit spacelike vector
$z^a=(l^a-n^a)/{\sqrt 2}$.  $\alpha = \alpha (x^\mu)$ and $\beta = \beta
(x^\mu)$ are real functions, $\beta$ is not identically zero, and $\gb \neq \pm
3 \ga$ for type I$_\bbR$.  It is possible to choose $\lambda=\pm 1$ and
$\tau=\pm 1$.

The first and second columns of the table respectively name the types by analogy
with the four-dimensional Petrov classification of the Weyl tensor and the Segre
classification of the traceless Ricci tensor.  The third column gives a
canonical form of the traceless Ricci tensor in terms of the null frame.  The
fourth column presents the traceless Ricci tensor in Jordan normal
form.\footnote{Any square matrix can be transformed, by a similarity
transformation, to Jordan normal form, which is block diagonal:
\bea
\mM = \begin{pmatrix} \mJ_1 & & \cr
         & \ddots & \cr
        && \mJ_p \end{pmatrix}\,,\nn
\eea
where each Jordan block $\mJ_i$ is of the form
\bea
\mJ_i = \begin{pmatrix} \lambda_i & 1 & & \cr
                       & \lambda_i &\ddots & \cr
	                     && \ddots & 1\cr
               &&& \lambda_i\end{pmatrix}\,,\nn
\eea
i.e.~$\lambda_i$ times an identity matrix plus 1's on the superdiagonal.}
The fifth column provides the basis for Jordan normal form.  The final column 
gives the minimal polynomial, which is the unique monic polynomial of 
minimum degree satisfied by $S{^a}{_b}$.

To precisely determine the Petrov--Segre type, we can find the Jordan 
normal form of $S{^a}{_b}$.  This is equivalent to finding the minimal 
polynomial of $S{^a}{_b}$.  In our three-dimensional situation, this is 
also equivalent to finding the eigenvalues of $S{^a}{_b}$, along with their 
algebraic and geometric multiplicities.  The Jordan normal form of $S{^a}{_b}$
is a convenient way of encoding any of this information.  In Segre
notation, the symbols 1, 2, 3 denote sizes of Jordan blocks.  Round brackets
group together Jordan blocks that correspond to the same eigenvalue.  Where
there is a comma, Jordan blocks before the comma correspond to spacelike
eigenvectors, and blocks after correspond to timelike eigenvectors.  This
refines the Petrov classification so that type D spacetimes are split according
to whether the one-dimensional eigenspace of $S{^\mu}{_\nu}$ is timelike, which
we denote as type D$_\ut$, or spacelike, which we denote as type D$_\us$. 
Jordan normal form by itself does not distinguish between types D$_\ut$ and
D$_\us$.  Although $S_{\mu \nu}$ is symmetric, $S{^\mu}{_\nu}$ is not symmetric
and can have complex eigenvalues.  We can refine the Petrov classification by
splitting Petrov type I spacetimes into type I$_\bbR$ (or $[11,1]$ in Segre
notation), which has three real distinct eigenvalues, and type I$_\bbC$ (or $[1
z \bar{z}]$ in Segre notation), which has one real and two complex conjugate
eigenvalues.  Each Jordan normal form has its own minimal polynomial.  Indeed, a
standard method of obtaining the Jordan normal form is to find the eigenvalues
from the characteristic equation, and then to find the minimal polynomial,
instead of computing the eigenvectors.

Note that determining the eigenvalues of $S{^a}{_b}$ and their
algebraic multiplicities is equivalent to finding the scalar
invariants\footnote{The normalisations 
$I = \tf{1}{2} S{^\mu}{_\nu} S{^\nu}{_\mu}$ and 
$J = \tf{1}{6} S{^\mu}{_\nu} S{^\nu}{_\gr} S{^\gr}{_\mu}$ are also frequently
used 
in the analogous four-dimensional literature.}
\ben
I := S{^\mu}{_\nu} S{^\nu}{_\mu} = \tr(\mS^2) \, , \qd J := S{^\mu}{_\nu}
S{^\nu}{_\gr} S{^\gr}{_\mu} = \tr(\mS^3) \, .
\la{IJ}
\een
Petrov--Segre types O, N and III satisfy the syzygies $I = J = 0$, types
D$_\ut$, D$_\us$ and II satisfy the syzygy $I^3 = 6 J^2 \neq 0$, and the most
general types I$_\bbR$ and I$_\bbC$ have respectively $I^3 > 6 J^2$ and $I^3 < 6 J^2$.

We record here the Petrov--Segre types of the previously known solutions of TMG: AdS$_3$ is type O, AdS pp-waves are type N, timelike-squashed AdS$_3$ is type D$_\ut$, spacelike-squashed AdS$_3$ is type D$_\us$, and triaxially squashed AdS$_3$ is type I$_\bbR$.  There are no solutions of types III, II and I$_\bbC$ in the literature.  Examples of type III and type II solutions are found in \cite{chposeII}.


\section{Squashed AdS$_3$}


We now prove that a Petrov--Segre type D solution of TMG must be biaxially squashed AdS$_3$.  First, we discuss properties of the squashed AdS$_3$ solutions.


\subsection{Biaxially squashed AdS$_3$}


   AdS$_3$ is a homogeneous spacetime that is in fact isomorphic to the
group manifold $\uSU (1,1)$ (or, locally, $\uSO (1,2)$).  It can be obtained
from $S^3$ by analytic continuation.  If we begin with the usual
left-invariant 1-forms of $\uSU (2)$, defined in terms of Euler angles
by
\be
\sigma_1=\cos\psi\, \ud \theta + \sin\psi\, \sin\theta\, \ud \phi\,,\quad
\sigma_2=-\sin\psi\, \ud \theta + \cos\psi\, \sin\theta\, \ud \phi\,,\quad
\sigma_3= \ud \psi+\cos\theta\, \ud \phi\, ,\label{su21forms}
\ee
we can consider ``triaxially squashed'' left-invariant metrics on $S^3$,
\be
\ds= \frac{1}{4} \sum_{i=1}^3 \lambda_i^2\, \sigma_i^2\,.
\ee
The round $S^3$, which is bi-invariant, is obtained by setting
$\lambda_1=\lambda_2=\lambda_3$.  If the squashing parameters $\gl_i$ are all taken to
be equal to 1, then this gives the unit-radius $S^3$.  Taking $\lambda_1=
\lambda_2$, and $\lambda_3=\lambda$, and for convenience setting the scale by
choosing $\lambda_1=\lambda_2=1$, we obtain the 1-parameter family of
biaxially squashed $S^3$ metrics
\ben
\ds= \ft14\lambda^2 (\ud \psi+\cos\theta\, \ud \phi)^2 + \ft14(\ud \theta^2 +
   \sin^2\theta\, \ud \phi^2)\, ,\label{biaxials3}
\een
which are sometimes known as Berger metrics.

   If we now make the analytic continuation and redefinition
\be
\theta\longrightarrow \ui\, \theta\,,\qquad \psi=\tau\,,
\ee
and then reverse the sign of the metric (thus giving, in the ``round'' case,
a negative cosmological constant), we obtain the 1-parameter family of
biaxial timelike-squashed AdS$_3$ metrics
\be
\ds = -\ft14\lambda^2 (\ud \tau +\cosh\theta\, \ud \phi)^2 + \ft14(\ud \theta^2
+
   \sinh^2\theta\, \ud \phi^2)\,.
\ee
We refer to this as a timelike squashing, since it is a timelike fibre
that is scaled relative to the size of the 2-dimensional base space (the
hyperbolic plane).

  There is a second type of biaxial squashing of AdS$_3$, which we refer
to as a spacelike squashing.  This is obtained by again starting from
(\ref{biaxials3}), but now making the analytic continuations and
redefinitions
\be
\theta= \ft12 \pi - \ui \rho\,,\qquad \phi=\tau\,,\qquad \psi=\ui z\,.
\ee
This gives, after reversing the sign of the original biaxially squashed
$S^3$ metric (\ref{biaxials3}), the 1-parameter family of biaxial 
spacelike-squashed AdS$_3$ metrics
\be
\ds = -\ft14 \cosh^2\rho\, \ud \tau^2
     +\ft14 \ud \rho^2 + \ft14\lambda^2\, (\ud z+\sinh\rho\, \ud \tau)^2 \,.
\ee
Here, the 2-dimensional base spacetime is AdS$_2$.

The timelike-squashed AdS$_3$ solution of TMG (with $\gL = - m^2$) has the
metric
\ben
\ds = \fr{9}{\mu^2 + 27 m^2}
\Big[ - \lambda^2(\ud \gt + \cosh \gq \, \ud \gf)^2 +
\ud \gq^2 + \sinh^2 \gq \, \ud \gf^2 \Big] \, ,
\la{tsquashedAdS}
\een
where the squashing parameter is defined as
\be
\lambda^2 = \fr{4 \mu^2}{\mu^2 + 27 m^2}
\la{squashpara}
\ee
If $\mu = \pm 3 m$, then the squashing parameter becomes 1, and the metric is
the standard ``round'' AdS$_3$.  The traceless Ricci tensor is
\ben
S_{\mu \nu} = (\tf{1}{9} \mu^2 - m^2) (g_{\mu \nu} + 3 k_\mu k_\nu) \, , \la{tsS}
\een
where $k$ is the unit timelike Killing vector
\ben
k^\mu \pd_\mu = \fr{\mu^2 + 27 m^2}{6 \mu} \fr{\pd}{\pd \gt} \, ,
\een
which has, by lowering the index, the associated 1-form
\ben
k_\mu \, \ud x^\mu =
- \fr{6 \mu}{\mu^2 + 27 m^2} (\ud \gt + \cosh \gq \, \ud \gf) \, .
\een
Therefore the metric has Petrov--Segre type D$_\ut$.  Examples of 
timelike-squashed AdS$_3$ appearing in the literature are given in Appendix 
\ref{A1}.

The spacelike-squashed AdS$_3$ solution of TMG has the metric
\ben
\ds = \fr{9}{\mu^2 + 27 m^2} \Big[- \cosh^2 \gr \,
\ud \gt^2 + \ud \gr^2 + \lambda^2 (\ud z + \sinh \gr \, \ud \gt)^2 \Big] \, ,
\la{ssquashedAdS}
\een
where again the squashing parameter is given by (\ref{squashpara}).  Again,
if $\mu = \pm 3 m$, then the squashing parameter becomes 1 and we have round
AdS$_3$.
The traceless Ricci tensor is
\ben
S_{\mu \nu} =
(\tf{1}{9} \mu^2 - m^2) (g_{\mu \nu} - 3 k_\mu k_\nu) \, , \la{ssS}
\een
where $k$ is the unit spacelike Killing vector
\ben
k^\mu \pd_\mu = \fr{\mu^2 + 27 m^2}{6 \mu} \fr{\pd}{\pd z} \, ,
\een
which has the associated 1-form
\ben
k_\mu \, \ud x^\mu = \fr{6 \mu}{\mu^2 + 27 m^2}
(\ud z + \sinh \gr \, \ud \gt) \, .
\een
Therefore the metric has Petrov--Segre type D$_\us$.  Examples of 
spacelike-squashed AdS$_3$ appearing in the literature are given in Appendix 
\ref{A2}.

For both the timelike- and spacelike-squashed AdS$_3$ solutions, the covariant
derivative of the Killing vector $k$ is
\ben
\na_\mu k_\nu = \tf{1}{3} \mu \gep_{\mu \nu \gr} k^\gr \, .
\la{nablak}
\een
In combination with the expressions for the traceless Ricci tensor in
terms of $k$ --- (\ref{tsS}) and (\ref{ssS}) --- it is straightforward to
check that the field equation (\ref{fieldeq2}) is solved.

We also have solutions for a positive cosmological constant, by taking $m \ra
\im m$.  Although there might appear to be a problem for $\gL = \tf{1}{27}
\mu^2$, for which the 2-dimensional base is flat, this is merely an artifact of
a bad coordinate choice.  For $\gL > \tf{1}{27} \mu^2$, we have squashing of de
Sitter spacetime.


\sse{From type D to squashed AdS$_3$}


Many of the solutions of TMG in the literature turn out locally to be biaxially
squashed AdS$_3$.  In light of this, it is useful to obtain a result that
enables us to determine whether or not a given solution is biaxially squashed
AdS$_3$, regardless of what coordinate system it is presented to us.

Our main result in this section will be to show that a Petrov--Segre type D solution of TMG is biaxially squashed AdS$_3$.  More specifically: a type D$_\ut$ solution, which has a traceless Ricci tensor of the form $S_{\mu \nu} = \ga (g_{\mu \nu} + 3 k_\mu k_\nu)$ with $k^\mu k_\mu = - 1$ and $\ga = \ga (x^\mu)$ a scalar function, is timelike-squashed AdS$_3$; a type D$_\us$ solution, which has $S_{\mu \nu} = \ga (g_{\mu \nu} - 3 k_\mu k_\nu)$ with $k^\mu k_\mu = 1$, is spacelike-squashed AdS$_3$.  We shall make
make use of this result in Appendices \ref{A1} and \ref{A2}, where we show that
various solutions in the literature are, by coordinate transformation, biaxially
squashed AdS$_3$ in disguise.  Analogously, all four-dimensional type D spacetimes that are vacuum are known \cite{kinner}, as are those that admit a cosmological constant \cite{dekamc, garsal}.

We begin by noting that Petrov--Segre type D$_\ut$ solution has a Ricci tensor of the form
\ben
R_{\mu \nu} = (p - 2 m^2) g_{\mu \nu} + 3 p k_\mu k_\nu \, , \label{RicciD}
\een
where $k$ is a timelike vector field normalized so that $k^\mu k_\mu =
-1$, and $p$ is a scalar function that is not identically zero.  Our first aim is to show that $p$ is constant and that $\na_\mu k_\nu = \tf{1}{3} \gep_{\mu \nu \gr} k^\gr$.  We then perform a dimensional reduction on the Killing vector $k$, showing that the 2-dimensional base space is Einstein, and then deducing that the spacetime is squashed AdS$_3$.

Using the vector field $k$, we perform a $2 + 1$ split of the
field equation components.  We define $h{^\mu}{_\nu} = \gd^\mu_\nu + k^\mu
k_\nu $, which satisfies $h{^\mu}{_\nu} k^\nu = 0$, to project out tensor
components orthogonal to $k$.  Contracting the field equation
(\ref{fieldeq2}) with $k^\mu k^\nu$, $k^\mu h{^\nu}{_\gs}$, $h{^\mu}{_\gr}
k^\nu$ and $h{^\mu}{_\gr} h{^\nu}{_\gs}$ gives a decomposition into four
simpler equations.  We hence obtain
\bea
&& \gep^{\mu \nu \gr} k_\mu \pd_\nu k_\gr = \tf{2}{3} \mu \, , \la{tt}
\w2
&& \gep^{\mu \nu \gr} k_\nu \pd_\gr p = 0 \, , \la{ts}
\w2
&& p h{^\gs}{_\gr} \gep^{\mu \nu \gr}
\pd_\mu k_\nu = \tf{2}{3} \gep^{\gs \mu \nu} k_\mu \pd_\nu p \, , \la{st}
\w2
&& p h_{\mu \nu} = \mu^{-1} (3 p \gep{_\mu}{^{\gr \gs}}
k_\gr \na_\gs k_\nu + h_{\mu \gr} h_{\nu \gs} \gep^{\gr \gs \gt} \pd_\gt p)
\, . \la{ss}
\eea
(Minus) the quantity on the left hand side of (\ref{tt}) is known as the scalar
twist of $k$, and is a constant here.  Substitution of (\ref{ts}) into
(\ref{st}) shows that $\gep^{\mu \nu \gr} \pd_\mu k_\nu$ is proportional to
$k^\gr$; the normalization is fixed by (\ref{tt}).  We hence obtain
\ben
\del_{[\mu} k_{\nu]} = \tf{1}{3} \mu \gep_{\mu \nu \gr} k^\gr \, , \la{st2}
\een
and then (\ref{tt}) is redundant.  Substituting (\ref{ts}) into
(\ref{ss}), we have
\ben
p h_{\mu \nu} = \mu^{-1} (3 p \gep{_\mu}{^{\gr \gs}}
k_\gr \na_\gs k_\nu + \gep_{\mu \nu \gr} \pd^\gr p ) \, . \la{ss2}
\een

We now show, by contradiction, that $p$ must be constant.  If $p$ is not
constant, 
then equation (\ref{ts})
implies that $k_\mu = q \pd_\mu p$ for some scalar function $q$,
i.e.~$k$ is hypersurface-orthogonal.  However, because $\mu \neq 0$, we
would have a contradiction with (\ref{tt}).  Therefore $p$ is constant.  Thus the entire content of the field equation reduces to $p$ being constant, (\ref{st2}), and, from (\ref{ss2}),
\ben
h_{\mu \nu} = 3 \mu^{-1} \gep{_\mu}{^{\gr \gs}} k_\gr \na_\gs k_\nu \, . \la{ss3}
\een

Our next step is to strengthen (\ref{st2}) to (\ref{nablak}).  The Bianchi identity $\na_{[ \mu} R_{\nu \gr ] \gs \gt} = 0$ is equivalent in three dimensions to the contracted Bianchi identity $\na^\mu G_{\mu \nu} = 0$, so in TMG is just $\nabla^\mu R_{\mu\nu}=0$, which then gives
\bea
&& \na_\mu k^\mu = 0 \, , \label{divk}
\w2
&& k^\mu \na_\mu k^\nu = 0 \, .
\eea
We now extend $k$ to an orthonormal frame, choosing the components of $k^a$ to be $(k^0, k^1, k^2) = (1, 0, 0)$.  Defining $K_{\mu \nu} := \na_\mu k_\nu$, from $\na_\mu (k^\nu k_\nu) = 0$ we have $K_{00} = K_{10} = K_{20} = 0$, and from the Bianchi identity we further have $K_{11} + K_{22} = 0$ and $K_{01} = K_{02} = 0$.  Then, from \eq{ss3}, we have $1 = h_{11} = - 3 \mu^{-1} K_{21}$, $1 = h_{22} = 3 \mu^{-1} K_{12}$, and $0 = h_{12} = - 3 \mu^{-1} K_{22}$.  The upshot is that the only non-vanishing components of $K_{ab}$ are $K_{12} = - K_{21} = \tf{1}{3} \mu$, and so
\ben
\na_\mu k_\nu = \tf{1}{3} \mu \gep_{\mu \nu \gr} k^\gr \, .
\la{nablak2}
\een
In particular, this relation implies that $k$ is a Killing vector.  Thus, we have reduced the content of the field equation and the Bianchi identity to $p$ being constant and (\ref{nablak2}).  The value of the constant $p$ is not fixed yet, since it is an overall factor in the field equation.

   With these results, we can now show that there is a metric that satisfies these properties, and that it must be squashed AdS$_3$.  It is convenient at this point to choose a coordinate system adapted to
the existence of the constant-length timelike Killing vector.  We may,
without loss of generality, write the metric as
\be
\ds = -(\ud t+ \cA)^2 + \ds_2 \, ,
\ee
where $\cA= \cA_i \, \ud x^i$ and $\ds_2 =h_{ij} \, \ud x^i \, \ud x^j$ depend
only on the two spatial coordinates $x^i$, with $i=1,2$.  The timelike Killing
vector, which is normalised to unit length, is $k^\mu \pd_\mu =\del/\del t$. 
The associated 1-form is given by $k_\mu \, \ud x^\mu =-(\ud t+\cA)$.  We choose
a sign so that the two- and three-dimensional metrics have respective volume
forms $\gep_2$ and $\gep_3$ related by
\ben
\gep_3 = - k \we \gep_2 \, ,
\la{e3e2}
\een
or in components $\gep_{tij} = \gep_{ij}$.  Noting that (\ref{st2})
can be written in terms of differential forms as
\be
\ud k=\ft23 \mu\, {*k}\,,
\la{dkeqn}
\ee
we see that
\be
\cF = \ft23\, \mu\, \ep_2\,,\label{cFep}
\ee
where $\cF:= \ud \cA$.

  The vielbein components $R_{ab}$ of the Ricci tensor for $\ds$
are related to the vielbein components $\bar R_{ij}$ of the metric 
$\ds_2 = \ud \bar s^2$ by
\be
R_{00} = \ft14 \cF_{ij} \cF^{ij}\,,\qquad
 R_{ij} = \bar R_{ij} + \ft12 \cF_{ik}\, \cF_{j}{}^k\,,\qquad
 R_{0i} = \ft12 \bar\nabla^j \cF_{ij}\,.
\ee
 From (\ref{RicciD}) and (\ref{cFep}) we therefore find that
\be
p= \ft19 \mu^2 -m^2\,,
\la{peqn}
\ee
and
\be
\bar R_{ij}= -\ft19 (\mu^2+ 27 m^2)\, \delta_{ij}\,.
\la{Rijeqn}
\ee
The base metric $\ds_2$ is therefore Einstein, with a negative cosmological
constant, i.e.~a hyperbolic space $\bbH^2$.

Locally, the base metric can be taken simply to be the standard metric on
the hyperbolic plane, given by
\be
\ds_2 = 9(\mu^2+27m^2)^{-1}\, (\ud \theta^2 + \sinh^2\theta\, \ud \phi^2)\,.
\ee
The volume form is $\ep_2= 9(\mu^2+27m^2)^{-1}\, \sinh\theta\, \ud \theta
\wedge \ud \phi$, and so from (\ref{cFep})
we may take the potential $\cA$ to be given by
\be
\cA =  6\mu\, (\mu^2+27m^2)^{-1}\, \cosh\theta\, \ud \phi\,.
\ee
Defining a new time coordinate $\tau = (\mu^2+27m^2) t / 6 \mu$, the
metric takes the form of (\ref{tsquashedAdS}), which is timelike-squashed
AdS$_3$.

We now turn to the spacelike case.  A Petrov--Segre type D$_\us$ solution
has a Ricci tensor of the form
\ben
R_{\mu \nu} =  (p - 2 m^2) g_{\mu \nu} - 3 p k_\mu k_\nu \, ,
\een
where $k$ is a spacelike vector field normalized so that $k^\mu k_\mu
= 1$, and $p$ is a scalar function that is not identically zero.  We
may perform the same analysis as in the case of type D$_\us$
solution; we can effectively replace $k \ra \im k$ in the equations
that result.  We again derive that $p$ must be constant and that $k$ is a Killing vector.  We may write the
metric in adapted coordinates as
\ben
\ds = \ud s_2^2 + (\ud y + \cA)^2 \, ,
\een
where $\cA = \cA_i \, \ud x^i$ and $\ds_2 = h_{ij} \, \ud x^i \, \ud
x^j$ depend only on the two spacetime coordinates $x^i$, with $i = 0,
1$.  The spacelike Killing vector is $k^\mu \pd_\mu = \pd / \pd y$, and has
associated 1-form $k_\mu \, \ud x^\mu = \ud y + \cA$.  Instead of choosing
(\ref{e3e2}), we choose
\ben
\gep_3 = k \wedge \gep_2 \, ,
\een
or in components $\gep_{ijy} = - \gep_{ij}$, and again have (\ref{dkeqn}),
(\ref{cFep}), (\ref{peqn}) and (\ref{Rijeqn}).  The base metric is therefore
anti-de Sitter spacetime AdS$_2$.

Locally, the base metric can be taken to be the AdS$_2$ metric in
global coordinates,
\ben
\ds_2 = 9(\mu^2+27m^2)^{-1} (- \cosh^2 \gr \, \ud \gt^2 + \ud \gr^2) \, .
\een
The volume form is $\ep_2= 9(\mu^2+27m^2)^{-1}\, \cosh \gr \, \ud \gt
\we \ud \gr$, and so from (\ref{cFep}) we may take the potential
$\cA$ to be given by
\be
\cA =  6\mu\, (\mu^2+27m^2)^{-1} \sinh \gr \, \ud \gt \, .
\ee
Defining a new spatial coordinate $z = (\mu^2+27m^2) y / 6 \mu$, the
metric takes the form of (\ref{ssquashedAdS}), which is spacelike-squashed
AdS$_3$.

We have taken $\gL \leq 0$ for definiteness, but entirely analogous results hold
for $\gL > 0$.  The only difference is that for $\gL = \tf{1}{27} \mu^2$ the
two-dimensional base space(time) is flat, and for $\gL > \tf{1}{27} \mu^2$ the
base space(time) has constant positive curvature.  In these cases, different
choices of explicit coordinates are required.


\section{AdS pp-waves}


We now complete the enumeration of solutions in the literature by reviewing the AdS pp-wave solutions of TMG.  Any solution of TMG that admits a null Killing vector field is (an AdS
generalization of) a pp-wave \ci{gipose}; we shall call such a solution an AdS
pp-wave.  Specifically, it was shown that in an adapted coordinate system for
which the null Killing vector is $k^\mu \pd_\mu=\partial/\partial v$, the
solution
for generic values $\mu \neq \pm m$ can take the form\footnote{The sign of
$\mu$ can be reversed by changing the orientation of the spacetime.}
\bea
\ds = \ud \rho^2 + 2 \ue^{2m\rho}\, \ud u\, \ud v + [\ue^{(m-\mu)\rho}\,
 f_1(u) + \ue^{2m\rho}\, f_2(u) + f_3(u) ]\, \ud u^2\,.\label{ppw1a}
\eea
In fact the functions $f_2(u)$ and $f_3(u)$ can be removed by means of
coordinate transformations, and so the AdS pp-wave solutions are characterised
by the single arbitrary function $f_1(u)$.  In the special cases $\mu=\pm m$,
the solutions, which can be obtained from (\ref{ppw1a}) by taking appropriate
limits, are
\bea
\mu=+m:&& \ds =  \ud \rho^2 + 2 \ue^{2m\rho}\, \ud u\, \ud v + [\rho\,
 f_1(u) + \ue^{2m\rho}\, f_2(u) + f_3(u) ]\, \ud u^2\,,\label{ppw1b}\\
\mu=-m:&& \ds =  \ud \rho^2 + 2 \ue^{2m\rho}\, \ud u \, \ud v + 
[\rho\,\ue^{2m\rho}\, 
 f_1(u) + \ue^{2m\rho}\, f_2(u) + f_3(u) ]\, \ud u^2\,.\label{ppw1c}
\eea
Again, the functions $f_2(u)$ and $f_3(u)$ can be removed by means of 
coordinate transformations.

  The traceless Ricci tensor takes the form
\ben
S_{\mu\nu}= c\,f_1(u)\,
\ue^{-(3m+\mu)\rho}\,  
k_\mu\, k_\nu \, ,
\label{Smunupp}
\een
where $c = \tf{1}{2} (m^2 - \mu^2)$ for $\mu \neq \pm m$ and $c = \mu$ for $\mu
= \pm m$.  The covariant derivative of $k$ is
\ben
\na_\mu k_\nu = - m \gep_{\mu \nu \gr} k^\gr \, .
\label{nablakpp}
\een
Combining these two expressions, using $\gep_{v u \gr} = + \sqrt{-g}$, provides
a check that we have a solution of TMG.  Examples of AdS pp-waves appearing in
the literature are given in Appendix \ref{A3}.


\se{Conclusion}


  In this paper, we have studied the algebraic classification of exact solutions in topologically massive
gravity, with a cosmological constant, described by the action (\ref{1.1}).  We have taken a first step towards classifying all possible 
solutions having a specific Petrov--Segre type.  Specifically, we 
have shown that a type D spacetime must be biaxial timelike-squashed or spacelike-squashed
AdS$_3$.

We have found that almost all the existing solutions are locally
equivalent, after coordinate transformation, to the biaxial timelike-squashed or
spacelike-squashed AdS$_3$, or to AdS pp-waves.  Thus locally, the previously
known solutions with non-vanishing cosmological constant are equivalent to
(\ref{tsquashedAdS}), (\ref{ssquashedAdS}) or (\ref{ppw1a}).  These are
respectively of Petrov--Segre types D$_\ut$, D$_\us$ and N.

In a subsequent paper \cite{chposeII}, we shall present a large class of solutions in TMG, which belong to the Kundt class of spacetimes.  These are generically of Petrov--Segre type II, but special cases are types III, N or D.  They provide the first known solutions of Petrov--Segre types II and III.


\section*{Acknowledgements}


This research has been supported in part by DOE Grant DE-FG03-95ER40917 and NSF Grant PHY-0555575. 

\appendix


\se{Literature Review}


Timelike- and spacelike-squashed AdS$_3$ and AdS pp-waves have been
independently rediscovered as solutions of topologically massive gravity several
times.  The literature is rather fragmented, using different coordinate systems
that are not obviously related.  We connect these fragments here by reviewing
how these three ubiquitous solutions appear in the literature.

Our aim is to be comprehensive; some of these connections have been known
previously.  However, we restrict ourselves to their introduction in the context of TMG.  Squashed AdS$_3$ has been studied in other contexts; see, for example, \ci{dulupo, bensan, anlipasost} for references.  Furthermore, we are only concerned with the local forms of the metrics here, not with their global
interpretations (for example, as black hole solutions obtained by quotienting by a discrete group).

For ease of comparison, we present the literature in our $- + +$ signature, and
we alter the notation of the original literature so that any cosmological
constant only appears via our parameter $m$, which has dimensions of mass, often
by replacing $\ell = \pm 1/m$, where $\ell$ has dimensions of length.  We
present the metrics with the same coordinates as the original literature,
although we may rearrange them.  Each metric is then transformed to a canonical
form, from which the conventions for $\mu^{-1} C_{\mu \nu}$ in the original
literature can be traced back.


\sse{Timelike-squashed AdS$_3$}\label{A2sec}
\label{A1}


To discover that the solutions here are (special cases of)
timelike-squashed AdS$_3$, we have applied the result obtained in
Section 3.  Namely, we found that these solutions all have Petrov--Segre type 
D$_\ut$, i.e.~the traceless Ricci tensor is of the form $S_{\mu \nu} = \ga
(g_{\mu \nu} + 3 k_\mu k_\nu)$ for some unit timelike vector $k$.

Recall from Section 3.2 that the metric of timelike-squashed AdS$_3$ can be
written in the form
\ben
\ds = - (\ud t + \cA)^2 + \ds_2 \, ,
\een
where $\ds_2$ is the metric of hyperbolic space $\bbH^2$ with squared radius
$L^2 = 9 / (\mu^2 + 27 m^2)$, which has volume-form $\gep_2$, and $\ud \cA =
\tf{2}{3} \mu \gep_2$.  It is helpful to review several coordinate systems. 
Two-dimensional hyperbolic space $\bbH^2$, with radius $L > 0$, is the upper
leaf $X^0 \geq L$ of the hyperboloid
\ben
- (X^0)^2 + (X^1)^2 + (X^2)^2 = - L^2
\een
in the flat three-dimensional spacetime with metric
\ben
\ds = - (\ud X^0)^2 + (\ud X^1)^2 + (\ud X^2)^2 \, .
\een
\begin{itemize}

\item The choice
\ben
X^0 = L \cosh \gq \, , \qd X^1 = L \sinh \gq \, \cos \gf \, , \qd X^2 = L \sinh
\gq \, \sin \gf
\een
gives $\bbH^2$ in polar coordinates:
\ben
\ds_2 = L^2 (\ud \gq^2 + \sinh ^2 \gq \, \ud \gf^2) \, .
\la{AdS2a}
\een

\item The choice
\ben
X^0 + X^2 = \fr{L^2}{y} \, , \qd X^0 - X^2 = \fr{x^2 + y^2}{y} \, , \qd X^1 =
\fr{L x}{y}
\een
gives $\bbH^2$ in Poincar\'{e} coordinates:
\ben
\ds_2 = \fr{L^2 (\ud x^2 + \ud y^2)}{y^2} \, .
\la{AdS2b}
\een

\item The choice
\ben
X^0 = L \cosh \gq \, \cosh \gf \, , \qd X^1 = L \sinh \gq \, , \qd X^2 = L \cosh
\gq \, \sinh \gf
\een
gives
\ben
\ds_2 = L^2 (\ud \gq^2 + \cosh^2 \gq \, \ud \gf^2) \, .
\la{AdS2c}
\een
\end{itemize}

We hereby express the timelike-squashed AdS$_3$ solution of TMG in several
different coordinate systems, with which we may compare the literature:
\bea
\ds & = & \fr{9}{\mu^2 + 27 m^2} \lt( - \fr{4 \mu^2}{\mu^2 + 27 m^2} 
(\ud \gt + \cosh \gq \, \ud \gf)^2 + \ud \gq^2 + \sinh^2 \gq \, \ud \gf^2 \rt) ,
\la{tAdS1} \\
\ds & = & \fr{9}{\mu^2 + 27 m^2} \bigg[ - \fr{4 \mu^2}{\mu^2 + 27 m^2} 
\lt( \ud \gt + \fr{\ud x}{y} \rt) ^2 + \fr{\ud x^2 + \ud y^2}{y^2} \bigg] ,
\la{tAdS2} \\
\ds & = & \fr{9}{\mu^2 + 27 m^2} \lt( - \fr{4 \mu^2}{\mu^2 + 27 m^2} 
(\ud \gt + \sinh \gq \, \ud \gf)^2 + \ud \gq^2 + \cosh^2 \gq \, \ud \gf^2 \rt) .
\la{tAdS3}
\eea
Relations between the coordinate systems for the $\bbH^2$ part of the metric
can 
be obtained via the expressions for $X^0$, $X^1$ and $X^2$.  The various $\gt$ 
coordinates above are distinct; relations between them can be obtained after 
relating the $\bbH^2$ coordinates.

The comparison with the literature is further helped by performing an
additional 
coordinate transformation on each of these three coordinate systems for
$\bbH^2$.  
If we make the coordinate changes $\cosh \gq = A r^2 + B$, $\gf \ra \gf / 2 A
L^2$ 
in (\ref{AdS2a}), $x = \gf / 2 A L^2$, $y = 1 / (A r^2 + B)$ in (\ref{AdS2b}) 
and $\sinh \gq = A r^2 + B$, $\gf \ra \gf / 2 A L^2$ in (\ref{AdS2c}), for 
some constants $A$ and $B$, then we obtain respectively
\bea
\ds_2 & = & \fr{4 L^2 r^2 \, \ud r^2}{(r^2 + B / A)^2 - 1 / A^2} + \fr{(r^2 + B /
A)^2 - 1 / A^2}{4 L^2} \, \ud \gf^2 \, , \\
\ds_2 & = & \fr{4 L^2 r^2 \, \ud r^2}{(r^2 + B / A)^2} + \fr{(r^2 + B / A)^2}{4
L^2} \, \ud \gf^2 \, , \\
\ds_2 & = & \fr{4 L^2 r^2 \, \ud r^2}{(r^2 + B / A)^2 + 1 / A^2} + \fr{(r^2 + B /
A)^2 + 1 / A^2}{4 L^2} \, \ud \gf^2 \, .
\eea
It follows that
\ben
\ds = - (\ud t - \tf{1}{3} \mu r^2 \ud \gf)^2 
+ \fr{r^2 \, \ud r^2}{\tf{1}{36} (\mu^2 + 27 m^2) r^4 + k_1 r^2 + k_0}
 + [\tf{1}{36} (\mu^2 + 27 m^2) r^4 + k_1 r^2 + k_0] \, \ud \gf^2 \, ,
\la{tAdS4}
\een
where $k_1$ and $k_0$ are arbitrary constants, is another way of 
writing the timelike-squashed AdS$_3$ solution.  The previous three 
separate coordinate systems respectively correspond to the discriminant 
$k_1^2 - \tf{1}{9} k_0 (\mu^2 + 27 m^2)$ being positive, zero or negative.


\subsubsection{With cosmological constant}



\pa{G\"{u}rses:}


G\"{u}rses \cite{Gurses:2008wu} considers solutions that are of
G\"{o}del type.  The most general solution is in his proposition 15, which is
defined by his equations (42), (43), (44), (46), and part of the unlabelled
equation following (39)\footnote{We refer to the numbering of the journal version, which differs slightly from that of the current arXiv version.}.  The solution is
\ben
\ds = - [\sr{a_0} \, \ud t + u_2 (r, \gq) \, \ud \gq + 
u_1 (r, \gq) \, \ud r]^2 + \fr{e_0^2 r^2 \psi (r)}{a_0} \, 
\ud \gq^2 + \fr{1}{\psi (r)} \, \ud r^2 \, ,
\la{gurses2a}
\een
where
\ben
\psi (r) = \frac{\mu^2 + 27 m^2}{36} r^2 + b_0 + \fr{b_1}{r^2} \, ,
\een
$u_1$ and $u_2$ are related by
\ben
\fr{\pd u_1}{\pd \gq} = \fr{\pd u_2}{\pd r} + \fr{2 \mu e_0}{3 \sr{a_0}} 
r \, ,
\la{gursesu}
\een
and $a_0$, $b_0$, $b_1$ and $e_0$ are arbitrary constants.  The general solution
of (\ref{gursesu}) can be expressed in terms of a potential function $U(r,
\gq)$, with
\ben
u_1 = \fr{\pd U}{\pd r} , \qd u_2 = \fr{\pd U}{\pd \gq} - 
\fr{\mu e_0}{3 \sr{a_0}} r^2 \, .
\een
Making the coordinate changes $t' = \sr{a_0} t + U$, $\gf = e_0 \gq / \sr{a_0}$,
followed by $t' \ra t$, the solution is (\ref{tAdS4}) with $k_1 = b_0$ and $k_0
= b_1$.

An earlier work of G\"{u}rses \cite{Gurses:1994} gives a solution in its
equation (5).  It is the above with, in the notation of (\ref{gurses2a}), $u_1 =
0$ and $u_2 = - (c_0 + \tf{1}{3} \mu e_0 r^2) / \sqrt{a_0}$, where $c_0$ is a
constant.


\pa{Nutku:}


Nutku \cite{nutku} generalized the solution of Vuorio
(\ref{vuoriosol}) to include a cosmological constant.  Solutions that
are timelike-squashed AdS$_3$ are given in several forms.

The solution in his equation (18), after making the coordinate relabelling $\psi
= \gt$, becomes (\ref{tAdS3}).

The solution in his equation (24), after making the coordinate relabelling $\gq
= \gf$, becomes (\ref{tAdS4}) with $k_1 = 1$ and $k_0 = 0$.

The black-hole solution in his equation (25) is
\bea
\ds & = & - \fr{2 J - M}{6} \lt( \ud t - \fr{2 \mu r^2 - 3 J / \mu}{2 J - M} 
\, \ud \gq \rt) ^2 + \fr{r^2 \, \ud r^2}{\tf{1}{36} (\mu^2 + 27 m^2) r^4 
- \tf{1}{6} M r^2  + \tf{1}{4} J^2 / \mu^2} \nnr
&& + \fr{6}{2 J - M} [\tf{1}{36} (\mu^2 + 27 m^2) r^4 - \tf{1}{6} M r^2 + 
\tf{1}{4} J^2 /\mu^2] \, \ud \gq^2 \, .
\eea
Making the coordinate changes
\ben
t' = \sqrt{\fr{2 J - M}{6}} \lt( t + \fr{3 J \gq}{\mu (2 J - M)} \rt) , 
\qd \gf = \sqrt{\fr{6}{2 J - M}}\, \gq \, ,
\een
followed by $t' \ra t$, the solution is (\ref{tAdS4}) with $k_1 = - \tf{1}{6} M$
and $k_0 = \tf{1}{4} J^2 / \mu^2$.


\pa{Cl\'{e}ment:}


Cl\'{e}ment \ci{clement} considers a Killing symmetry reduction procedure
to obtain stationary rotationally symmetric solutions.  The solution in
his equation (18) is
\bea
\ds & = & - \lt[ \sr{2 a} \, \ud t + \lt( \fr{3 b}{\mu} - 
\fr{2 \mu \gr}{3} \rt) \fr{\ud \gq}{\sr{2 a}} \rt] ^2 + \fr{\ud \gr^2}{\tf{1}{9}
(\mu^2 + 27 m^2) \gr^2 + 
4 (a - b) \gr + 9 b^2 / \mu^2} \nnr
&& + 
[\tf{1}{9} (\mu^2 + 27 m^2) \gr^2 + 4 (a - b) \gr + 9 b^2 / \mu^2] 
\fr{\ud \gq^2}{2 a} \, .
\eea
Making the coordinate changes
\ben
t' = \sr{2a} t + \fr{3 b \gq}{\mu \sr{2 a}} \, , \qd r^2 = \gr \, , \qd \phi =
\sqrt{\fr{2}{a}} \gq \, ,
\een
followed by $t' \ra t$, the solution is (\ref{tAdS4}) with $k_1 = a - b$ and
$k_0 = \tf{9}{4} b^2 / \mu^2$.


\pa{Anninos--Li--Padi--Song--Strominger:}


Anninos, Li, Padi, Song and Strominger \ci{anlipasost} consider ``warped''
AdS$_3$ black hole solutions of TMG.  Making the coordinate relabellings $\gs =
\gq$, $u = \gf$, the solution in their equation (3.4) becomes (\ref{tAdS3}).


\ssse{Without cosmological constant}



\pa{Vuorio:}


Vuorio \cite{vuorio} finds a stationary rotationally symmetric solution.  The
solution of his equation (2.21) is
\ben
\ds = \fr{9}{\mu^2} [ - (\ud t + 2 \, \ud \gq - 2 \cosh \gs \, \ud \gq) ^2 
+ ( \ud \gs^2 + \sinh^2 \gs \, \ud \gq^2) ] \, . \la{vuoriosol}
\een
Making the coordinate changes $t \ra 2 (\gt + \gf)$, $\gq \ra - \gf$, $\gs \ra
\gq$, the solution is (\ref{tAdS1}) with $m = 0$.


\pa{Percacci--Sodano--Vuorio:}


Percacci, Sodano and Vuorio \cite{Percacci:1986ja} consider stationary solutions
for which the timelike Killing vector has a constant scalar twist.  The solution in their equation (3.20) is
\ben
\ds = - 3 [\ud x^2 + \tf{2}{3} \exp(\tf{1}{3} \mu x^1) \, \ud x^0]^2 + 
(\ud x^1)^2 + \tf{1}{3} \exp(\tf{2}{3} \mu x^1) \, (\ud x^0)^2 \, .
\een
Making the coordinate changes $\gt = \mu x^2 / (2 \sr{3})$, $x = \mu x^0
/ (3 \sr{3})$, $y = \exp(- \tf{1}{3} \mu x^1)$, the solution is
(\ref{tAdS2}) with $m = 0$.


\pa{Nutku--Baekler and Ortiz:}


Nutku and Baekler \ci{nutbak} and Ortiz \ci{ortiz} consider solutions formed
from left-invariant 1-forms of Bianchi spaces.  Some of these solutions are
timelike-squashed AdS$_3$.


\subparagraph{Bianchi VIII:}


The solution in equation (4.1) of \ci{nutbak} and in \ci{ortiz}, where it is
called type (a) with $a = 0$, is the triaxially squashed AdS$_3$
\bea
\ds & = & - \gl_0^2 \gs_3^2 + \gl_1^2 \gs_1^2 + \gl_2^2 \gs_2^2 \nnr
& = & - \gl _0^2 (\ud \psi + \sinh \gq \, \ud \gf)^2 
+ \gl_1^2 (- \sin \psi \, \ud \gq + \cos \psi \, \cosh \gq \, \ud \gf)^2 \nnr
&& + \gl_2^2 (\cos \psi \, \ud \gq + \sin \psi \, \cosh \gq \, \ud \gf)^2 \, ,
\eea
where $\gl_0$, $\gl_1$ and $\gl_2$ are constants that satisfy, after choosing
appropriate signs, $\gl_0 + \gl_1 + \gl_2 = 0$, and $\gs_i$ are Bianchi VIII
left-invariant 1-forms that satisfy $\ud \gs_1 = \gs_2 \wedge \gs_3$, $\ud \gs_2
= \gs_3 \wedge \gs_1$ and $\ud \gs_3 = - \gs_1 \wedge \gs_2$.  It is a solution
of TMG with $\mu = \pm (\gl_0^2 + \gl_1^2 + \gl_2^2) / \gl_0 \gl_1
\gl_2$,\footnote{We correct a factor of 2 in $\mu$.} the sign depending on the
orientation.  If $\gl_1 = \gl_2$, then making the coordinate relabelling $\gt =
\psi$ gives timelike-squashed AdS$_3$ in the form (\ref{tAdS3}) with $m = 0$.


\subparagraph{Bianchi III:}


One of the solutions, by choice of signs, in equation (4.10)\footnote{We correct
a typographical error in $g_{\gq \gq}$.} of \ci{nutbak} and in \ci{ortiz}, where
it is called type (a) with $a \neq 0$, is
\bea
\ds & = & \gl_1^2 [- (2 + \sr{3})^2 \gs_1^2 + \gs_2^2] + \fr{36}{\mu^2} \gs_3^2
\nnr
& = & \gl_1^2 \ue^{2 \gq} [ - (2 + \sr{3})^2 (\cosh \gq \, \ud x + \sinh \gq \,
\ud y)^2 + (\sinh \gq \, \ud x + \cosh \gq \, \ud y)^2 ] + \fr{36}{\mu^2} \, \ud
\gq^2 \nnr
& = & -\fr{3 + 2 \sr{3}}{6} \gl_1^2 [\sr{3} (\ud x - \ud y) + 2 \ue^{2 \gq} (\ud
x + \ud y)]^2 + \fr{36}{\mu^2} \, \ud \gq^2 + \fr{3 + 2 \sr{3}}{6} \gl_1^2
\ue^{4 \gq} (\ud x + \ud y)^2 \, , \nnr
\eea
where $\gs_i$ are Bianchi III\footnote{\cite{nutbak} refers to these as being
Bianchi VI, however they are the Bianchi III limit of Bianchi VI.}
left-invariant 1-forms that satisfy $\ud \gs_1 = \gs_3 \wedge \gs_2$, $\ud \gs_2
= \gs_3 \wedge \gs_1$ and $\ud \gs_3 = 0$.  Making the coordinate changes
\ben
\gt = \fr{\mu \sr{3 + 2 \sr{3}} \gl_1 (x - y)}{6 \sr{2}} \, , \qd x' = \fr{\mu
\sr{3 + 2 \sr{3}} \gl_1 (x + y)}{3 \sr{6}} \, , \qd y' = \ue^{- 2 \gq} \, ,
\een
followed by $x' \ra x$, $y' \ra y$, the solution is (\ref{tAdS2}) with $m = 0$.


\pa{Cl\'{e}ment}


Cl\'{e}ment \cite{clement92} (see also \cite{clement92a}) considers stationary
rotationally symmetric solutions.

The solution in his equation (4.4) is
\ben
\ds = - [\ud t + 2 c \, \ud \gq - 2 c \cosh (\tf{1}{3} \mu r) \, \ud \gq]^2 +
\ud r^2 + c^2 \sinh^2 (\tf{1}{3} \mu r) \, \ud \gq^2 \, .
\een
Making the coordinate changes $\gt = \tf{1}{6} \mu (t + 2 c \gq)$, $\gq' =
\tf{1}{3} \mu r$, $\gf = - \tf{1}{3} \mu c \gq$, followed by $\gq' \ra \gq$, the
solution is (\ref{tAdS1}) with $m = 0$.

The solution in his equation (4.5) is
\ben
\ds = - [\ud t - 2 c \sinh (\tf{1}{3} \mu r) \, \ud \gq]^2 + \ud r^2 + c^2
\cosh^2 (\tf{1}{3} \mu r) \, \ud \gq^2 \, .
\een
Making the coordinate changes $\gt = \tf{1}{6} \mu t$, $\gq' = \tf{1}{3} \mu r$,
$\gf = - \tf{1}{3} \mu c \gq$, followed by $\gq' \ra \gq$, the solution is
(\ref{tAdS3}) with $m = 0$.

The solutions in his equation (4.7) have various choices of signs.  Two choices
of signs give
\ben
\ds = - \fr{c}{d} (\ud t \mp 2 d \ue^{\pm \mu r / 3} \, \ud \gq)^2 + \ud r^2 + c
d \ue^{\pm 2 \mu r / 3} \, \ud \gq^2 \, .
\een
Making the coordinate changes $\gt = \sr{c/d} \, t$, $x = \mp \tf{1}{3} \mu
\sr{c d} \, \gq$, $y = \ue^{\mp \mu r / 3}$, the solution is
(\ref{tAdS2}) with $m = 0$.  Another two choices of signs give
\ben
\ds = - 3 c d \lt( \ud \gq \pm \fr{2}{3 d} \ue^{\pm \mu r / 3} \, \ud t \rt) ^2
+ \ud r^2 + \fr{c}{3 d} \ue^{\pm 2 \mu r / 3} \, \ud t^2 \, .
\een
Making the coordinate changes $\gt = \sr{3 c d} \, \gq$, $x = \pm \mu \sr{c / 27
d} \, t$, $y = \ue^{\mp \mu r / 3}$, the solution is
(\ref{tAdS2}) with $m = 0$.


\sse{Spacelike-squashed AdS$_3$}
\label{A2}


To discover that the solutions here are (special cases of) spacelike-squashed
AdS$_3$, we have applied the result obtained in Section 3.  Namely, we found
that these solutions all have Petrov--Segre type D$_\us$, i.e.~$S_{\mu \nu} =
\ga (g_{\mu \nu} - 3 k_\mu k_\nu)$ for some unit spacelike vector $k$.

Recall from Section 3.2 that the metric of spacelike-squashed AdS$_3$ can be
written in the form
\ben
\ds = \ds_2 + (\ud y + \cA)^2 \, ,
\een
where $\ds_2$ is the metric of AdS$_2$ with squared radius $L^2 = 9 / (\mu^2 +
27 m^2)$, which has volume form $\gep_2$, and $\ud \cA = \tf{2}{3} \mu \gep_2$. 
It is helpful to review several coordinate systems.  Two-dimensional anti-de
Sitter spacetime AdS$_2$, with AdS radius $L > 0$, is the hyperboloid
\ben
- (X^0)^2 + (X^1)^2 - (X^2)^2 = - L^2
\een
in the flat three-dimensional spacetime with metric
\ben
\ds = - (\ud X^0)^2 + (\ud X^1)^2 - (\ud X^2)^2 \, .
\een
\begin{itemize}

\item The choice
\ben
X^0 = L \cosh \gr \, \cos \gt \, , \qd X^1 = L \sinh \gr \, , \qd X^2 = L \cosh
\gr \, \sin \gt
\een
gives AdS$_2$ in global coordinates:
\ben
\ds_2 = L^2 (- \cosh^2 \gr \, \ud \gt^2 + \ud \gr^2) \, .
\een

\item The choice
\ben
X^0 = \fr{L t}{\gr} \, , \qd X^1 + X^2 = \fr{L^2}{\gr} , \qd X^2 - X^1 = \gr -
\fr{t^2}{\gr}
\een
gives AdS$_2$ in conformally flat coordinates that cover the Poincar\'{e} patch:
\ben
\ds_2 = \fr{L^2 (- \ud t^2 + \ud \gr^2)}{\gr^2} \, .
\een

\item The choice
\ben
X^0 = L \cosh \gr \, , \qd X^1 = L \sinh \gr \, \cosh \gt \, , \qd X^2 = L \sinh
\gr \, \sinh \gt
\een
gives AdS$_2$ in coordinates that cover only $X^0 \geq L$:
\ben
\ds_2 = L^2 (- \sinh^2 \gr \, \ud \gt^2 + \ud \gr^2) \, .
\een

\item The choice
\ben
X^0 = L \sin \gt \, , \qd X^1 = L \cos \gt \, \sinh \gf \, , \qd X^2 = L \cos \gt
\, \cosh \gf
\een
gives AdS$_2$ in coordinates that cover only $-L \leq X^0 \leq L$:
\ben
\ds_2 = L^2 (- \ud \gt^2 + \cos^2 \gt \, \ud \gf^2) \, .
\een
\end{itemize}

We hereby express the spacelike-squashed AdS$_3$ solution of TMG in several
different coordinate systems, with which we may compare the literature:
\bea
\ds & = & \fr{9}{\mu^2 + 27 m^2} \lt( -\cosh^2 \gr \, \ud \gt^2 + \ud \gr^2 
+ \fr{4 \mu^2}{\mu^2 + 27 m^2} (\ud z + \sinh \gr \, \ud \gt)^2 \rt) ,
\la{sAdS1} \\
\ds & = & \fr{9}{\mu^2 + 27 m^2} \bigg[ \fr{- \ud t^2 + \ud x^2}{x^2} + 
\fr{4 \mu^2}{\mu^2 + 27 m^2} \lt( \ud z + \fr{\ud t}{x} \rt) ^2 \bigg] ,
\la{sAdS2} \\
\ds & = & \fr{9}{\mu^2 + 27 m^2} \lt( -\sinh^2 \gr \, \ud \gt^2 + \ud \gr^2 
+ \fr{4 \mu^2}{\mu^2 + 27 m^2} (\ud z + \cosh \gr \, \ud \gt)^2 \rt) ,
\la{sAdS3} \\
\ds & = & \fr{9}{\mu^2 + 27 m^2} \lt( - \ud \gt^2 + \cos^2 \gt \, \ud \gf^2
+ \fr{4 \mu^2}{\mu^2 + 27 m^2} (\ud z + \sin \gt \, \ud \gf)^2 \rt) . \la{sAdS4}
\eea
Relations between the coordinate systems for the AdS$_2$ part of the metric can
be obtained via the expressions for $X^0$, $X^1$ and $X^2$.  The various $z$
coordinates above are distinct; relations between them can be obtained after
relating the AdS$_2$ coordinates.


\ssse{With cosmological constant}



\pa{Bouchareb--Cl\'{e}ment:}


Bouchareb and Cl\'{e}ment \ci{boucle} construct black hole solutions.
The solution in their equation (4.1) is
\bea
\ds & = & - \fr{\mu^2 + 27 m^2}{3 (\mu^2 - 9 m^2)} (\gr^2 - \gr_0^2) 
\, \ud \gvf^2 + \fr{9}{\mu^2 + 27 m^2} \, \fr{\ud \gr^2}{\gr^2 - \gr_0^2} \nnr
&& + \fr{3 (\mu^2 - 9 m^2)}{4 \mu^2} \lt[ \ud t - 
\lt( \fr{4 \mu^2 \gr}{3 (\mu^2 - 9 m^2)} + \gw \rt) \ud \gvf \rt] ^2 .
\eea
Making the coordinate changes
\ben
\gt = \fr{\gr_0 (\mu^2 + 27 m^2) \gvf}{3 \sr{3} \sr{\mu^2 - 9 m^2}} , 
\qd \cosh \gr' = \fr{\gr}{\gr_0} , \qd z = \fr{\sr{\mu^2 - 9 m^2} 
(\mu^2 + 27 m^2) (\gw \gvf - t)}{4 \sr{3} \mu^2} \, ,
\een
and then $\gr' \ra \gr$, the solution is (\ref{sAdS3}).


\pa{Anninos--Li--Padi--Song--Strominger:}


Anninos, Li, Padi, Song and Strominger \ci{anlipasost} consider ``warped''
AdS$_3$ black hole solutions of TMG.  After making the coordinate relabellings
$\gs = \gr$, $u = z$, the solution in their equation (3.3) becomes
(\ref{sAdS1}).


\ssse{Without cosmological constant}



\pa{Hall--Morgan--Perj\'{e}s:}


Hall, Morgan and Perj\'{e}s \ci{hamope} look for solutions of TMG with a
particular (Petrov)--Segre type.  The solution in their equation (61) is
\ben
\ds = 2 \, \ud u \, \ud v - \tf{1}{9} \mu^2 v^2 \, \ud u^2 + 
(\ud y - \tf{2}{3} \mu v \, \ud u)^2 + 2 f(u) v \, \ud u^2 \, .
\een
Taking $F(u) = \int \ud u \, f(u)$ and making the coordinate changes
$u' = \int \ud u \, \ue^{- F(u)}$, $v' = \ue^{F(u)} v$, we see that
the function $f$ is redundant.  After taking $f = 0$, we make the
coordinate changes $\hat{u} = \tf{1}{18} \mu^2 u$, $\hat{v} = 1/v +
\tf{1}{18} \mu^2 u$, so the solution is
\ben
\ds = - \fr{36 \, \ud \hat{u} \, \ud \hat{v}}{\mu^2 
(\hat{u} - \hat{v})^2} + \lt( \ud y + \fr{12}{\mu} 
\fr{\ud \hat{u}}{\hat{u} - \hat{v}} \rt) ^2 \, .
\een
Then making the coordinate changes $t = \tf{1}{2} (\hat{u} +
\hat{v})$, $x = \tf{1}{2} (\hat{v} - \hat{u})$, $z = - \tf{1}{6} \mu y
- \log[\tf{1}{2} (\hat{v} - \hat{u})]$, the solution is
(\ref{sAdS2}) with $m = 0$.


\pa{Nutku--Baekler and Ortiz:}


Nutku and Baekler \ci{nutbak} and Ortiz \cite{ortiz} consider solutions formed
from left-invariant 1-forms of Bianchi spaces.  Some of these solutions are
spacelike-squashed AdS$_3$.


\subparagraph{Bianchi VIII:}


The triaxially squashed AdS$_3$ solution of \ci{nutbak} and \ci{ortiz} is
\ben
\ds = - \gl_0^2 \gs_3^2 + \gl_1^2 \gs_1^2 + \gl_2^2 \gs_2^2 \, ,
\een
where $\gl_0$, $\gl_1$ and $\gl_2$ are constants that satisfy, after choosing
appropriate signs, $\gl_0 +\gl_1 + \gl_2 = 0$, and $\gs_i$ are Bianchi VIII
left-invariant 1-forms that satisfy $\ud \gs_1 = \gs_2 \wedge \gs_3$, $\ud \gs_2
= \gs_3 \wedge \gs_1$ and $\ud \gs_3 = - \gs_1 \wedge \gs_2$.  It is a solution
of TMG with $\mu = \pm (\gl_0^2 + \gl_1^2 + \gl_2^2) / \gl_0 \gl_1
\gl_2$,\footnote{We correct a factor of 2 in $\mu$.} the sign depending on the
orientation.

The solution is presented in equation (4.6) of \ci{nutbak} as
\bea
\ds & = & - \gl_0^2 (\cosh \psi \, \ud \gq + \sinh \psi 
\, \cos \gq \, \ud \gf)^2 + \gl_1^2 (\sinh \psi \, \ud \gq + 
\cosh \psi \, \cos \gq \, \ud \gf)^2 \nnr
&& + \gl_2^2 (\ud \psi + \sin \gq \, \ud \gf)^2 \, ,
\eea
If $\gl_0 = \gl_1$, then making the coordinate relabellings $\gt = \gq$, $z =
\psi$ gives spacelike-squashed AdS$_3$ in the form (\ref{sAdS4}) with $m = 0$.

The solution is presented in equation (4.8) of \ci{nutbak} as\footnote{We
correct a typographical error in the $\gl_1^2$ term.}
\bea
\ds & = & - \gl_0^2 (- \sinh \psi \, \ud \gq + \cosh \psi \, 
\cosh \gq \, \ud \gf)^2 + \gl_1^2 (\cosh \psi \, \ud \gq - 
\sinh \psi \, \cosh \gq \, \ud \gf)^2 \nnr
&& + \gl_2^2 (\ud \psi + \sinh \gq \, \ud \gf)^2 \, ,
\eea
If $\gl_0 = \gl_1$, then making the coordinate relabellings $\gt = \gf$, $\gr =
\gq$, $z = \psi$ gives spacelike-squashed AdS$_3$ in the form (\ref{sAdS1}) with
$m = 0$.


\subparagraph{Bianchi III:}


One of the solutions, by choice of signs, in equation (4.10)\footnote{We correct
a typographical error in $g_{\gq \gq}$.}	of \ci{nutbak} and in
\ci{ortiz}, where it is called type (a) with $a \neq 0$, is
\bea
\ds & = & \gl_1^2 [- (2 - \sr{3})^2 \gs_1^2 + \gs_2^2] + \fr{36}{\mu^2} \gs_3^2
\nnr
& = & \gl_1^2 \ue^{2 \gq} [ - (2 - \sr{3})^2 (\cosh \gq \, \ud x + \sinh \gq \,
\ud y)^2 + (\sinh \gq \, \ud x + \cosh \gq \, \ud y)^2 ] + \fr{36}{\mu^2} \, \ud
\gq^2 \nnr
& = & - \fr{2 \sr{3} - 3}{6} \gl_1^2 \ue^{4 \gq} (\ud x + \ud y)^2 +
\fr{36}{\mu^2} \, \ud \gq^2 + \fr{2 \sr{3} - 3}{6} \gl_1^2 [\sr{3} (\ud x - \ud
y) - 2 \ue^{2 \gq} (\ud x + \ud y)]^2 \, . \nnr
\eea
where $\gs_i$ are Bianchi III\footnote{\cite{nutbak} refers to these as being
Bianchi VI, however they are the Bianchi III limit of Bianchi VI.}
left-invariant 1-forms that satisfy $\ud \gs_1 = \gs_3 \wedge \gs_2$, $\ud \gs_2
= \gs_3 \wedge \gs_1$ and $\ud \gs_3 = 0$.  Making the coordinate changes
\ben
t = \fr{\mu \sr{2 \sr{3} - 3} \gl_1 (x + y)}{3 \sr{6}} \, , \qd x' = \ue^{- 2
\gq} \, , \qd z = \fr{\mu \sr{2 \sr{3} - 3} \gl_1 (x - y)}{6 \sr{2}} \, ,
\een
followed by $x' \ra x$, the solution is (\ref{sAdS2}) with $m = 0$.


\pa{Ait Moussa--Cl\'{e}ment--Leygnac:}


Ait Moussa, Cl\'{e}ment and Leygnac \ci{aiclle} obtain black hole
solutions by analytically continuing the solution of Vuorio
(\ref{vuoriosol}).  The solution in their equation (4) is
\ben
\ds = \fr{9}{\mu^2} \lt( - \fr{\gr^2 - \gr_0^2}{3} \, 
\ud \gvf^2 + \fr{1}{\gr^2 - \gr_0^2} \, \ud \gr^2 + 
3 [\ud t - (\tf{2}{3} \gr + \gw) \, \ud \gvf]^2 \rt) \, .
\een
Making the coordinate changes $\gt = \gr_0 \gvf / \sr{3}$, $\cosh \gr'
= \gr / \gr_0$, $z = \sr{3} (\gw \gvf - t) / 2$, and then $\gr' \ra \gr$, the
solution is (\ref{sAdS3}) with $m = 0$.


\sse{AdS pp-waves}
\label{A3}


To discover that the solutions here are (special cases of) the AdS pp-wave, 
we have applied the results of \ci{gipose}.  Namely, we have found that these
solutions have Petrov--Segre type N, i.e.~$S_{\mu \nu} = k_\mu k_\nu$ for some
null vector $k$, and furthermore $k$ is proportional to a Killing vector.  The
general pp-wave solutions with a non-vanishing cosmological constant are given
by (\ref{ppw1a}), (\ref{ppw1b}) and (\ref{ppw1c}); we deal later with the zero
cosmological constant limit.  Although the functions $f_2(u)$ and $f_3(u)$ can
be removed by coordinate transformations, it is convenient to include them when
making comparisons with the literature.


\ssse{With cosmological constant}



\pa{Nutku:}


Nutku \ci{nutku} considers several solutions with a cosmological constant.  The
solution in his equations (16) and (17) is\footnote{We correct a typographical
error in the $uu$ component.}
\ben
\ds = \fr{\ud x^2 - 2 \, \ud u \, \ud v + 2 \{ - x \ddot{h}(u) / m 
- c [m x + h(u)]^{1 + \mu / m} \} \, \ud u^2}{[m x + h(u)]^2} \, ,
\een
where $h(u)$ is an arbitrary function and $c$ is a constant.  Making the
coordinate changes $\gr = - \log (m x + h) / m$, $v' = - v + \dot{h} (m x + h) /
m^2$, and then $v' \ra v$, the solution is (\ref{ppw1a}) with $f_1 = - c$,
$f_2 = (2 h \ddot{h} + \dot{h}^2) / m^2$, $f_3 = 0$.


\pa{Cl\'{e}ment:}


Cl\'{e}ment \ci{clement} considers a Killing symmetry reduction procedure to
obtain stationary rotationally symmetric solutions.  The solution in his
equation (21) is\footnote{We have absorbed a sign ambiguity by taking $m = \pm 1
/ l$.}
\ben
\ds = \fr{\ud \gr^2}{4 m^2 \gr^2} + 2 m^2 \gr 
\lt( \ud t^2 - \fr{\ud \gq^2}{m^2} \rt) + \fr{M}{2} 
(1 + c \gr^{(1 - \mu / m)/2}) \lt( \ud t - \fr{\ud \gq}{m} \rt) ^2 \, ,
\een
where $M$ and $c$ are constants.  Making the coordinate changes $\gr' = (\log
\gr) / 2 m$, $u = t - \gq / m$, $v = (t + \gq / m) / 2 m^2$, and then $\gr' \ra
\gr$, the solution is (\ref{ppw1a}), with $f_1 = \tf{1}{2} M c$, $f_2 = 0$, $f_3
= \tf{1}{2} M$.


\pa{Ay\'{o}n-Beato--Hassa\"{\i}ne:}


Ay\'{o}n-Beato and Hassa\"{\i}ne \cite{ayohas3} obtain AdS pp-waves by using the
general AdS pp-wave ansatz.  The solutions in their equations (A3), (A5) and
(A4) are respectively
\bea
\ds & = & \fr{1}{m^2 y^2} [\ud y^2 - 2 \, \ud u \, \ud v - (m y)^{1 + \mu / m}
F_1 (u) \, \ud u^2] \, , \\
\ds & = & \fr{1}{m^2 y^2} [\ud y^2 - 2 \, \ud u \, \ud v - y^2 \log(- m y) F_1
(u) \, \ud u^2] \, , \\
\ds & = & \fr{1}{m^2 y^2} [\ud y^2 - 2 \, \ud u \, \ud v - \log(- m y) F_1 (u)
\, \ud u^2] \, ,
\eea
where $F_1 (u)$ is an arbitrary function, have respectively $\mu \neq \pm m$,
$\mu = m$ and $\mu = - m$.  Making the coordinate changes $\gr = - [\log
(my)]/m$, $v \ra - v$, the solutions are respectively (\ref{ppw1a}),
(\ref{ppw1b}) and (\ref{ppw1c}), with $f_1 = - F_1$, $f_2 = 0$, $f_3 = 0$.

They had previously found \cite{ayohas1} AdS pp-waves by considering a
correspondence between Cotton gravity with a conformally coupled scalar field
and TMG \ci{dejapi}.  The general $\mu = - m$ solution cannot be obtained
through this correspondence.


\pa{\"{O}lmez--Sar\i o\u{g}lu--Tekin and Dereli--Sar\i o\u{g}lu:}


\"{O}lmez, Sar\i o\u{g}lu and Tekin \cite{olsate} consider supersymmetric
solutions.  The solution in their equation (9) is\footnote{We have taken $m =
\mp 1/l$.}
\ben
\ds = \ud \gr^2 + 2 \ue^{\mp 2 m \gr} \, \ud u \, \ud v + [\gb_2 (v) \ue^{\mp (m
+ \mu) \gr} + \gb_1 (v) \ue^{\mp 2 m \gr} + \gb_0 (v)] \, \ud v^2 \, ,
\een
where $\gb_0 (v)$, $\gb_1 (v)$ and $\gb_2 (v)$ are arbitrary functions.  Making
the coordinate changes $u \leri v$, $\gr \ra \mp \gr$, the solution is
(\ref{ppw1a}) with $f_1 = \gb_2$, $f_2 = \gb_1$, $f_3 = \gb_0$.

Previously, Dereli and Sar\i o\u{g}lu \ci{dersar} considered supersymmetric
solutions, but with $\gb_0$, $\gb_1$ and $\gb_2$ constants.  The solution in
their equations (34) to (36) is a special case of the above.


\pa{Anninos--Li--Padi--Song--Strominger:}


Anninos, Li, Padi, Song and Strominger \ci{anlipasost} consider ``warped''
AdS$_3$ black hole solutions of TMG.  The solution, for $\mu = - 3 m$, in their
equation (3.7) is
\ben
\ds = \fr{1}{m^2} \lt( \fr{\ud u^2 + \ud x^+ \, \ud x^-}{u^2} 
+ \fr{(\ud x^-)^2}{u^4} \rt) .
\een
Making the coordinate changes $\gr = - \log (m u) / m$, $u' = m^2
x^-$, $v = x^+ / 2 m^2$, and then $u' \ra u$, the solution is
(\ref{ppw1a}) with $\mu = - 3 m$, $f_1 = 1$, $f_2 = 0$, $f_3 = 0$.


\pa{Carlip--Deser--Waldron--Wise:}


Carlip, Deser, Waldron and Wise \ci{cadewawi} write down an AdS
pp-wave solution.  The solution, for $\mu \neq \pm m$, in their
equation (21) is
\ben
\ds = \fr{\ud z^2 + 2 \, \ud x^+ \, \ud x^- + 
2 (m z)^{1 + \mu / m} h(x^+) \, (\ud x^+)^2}{m^2 z^2} \, ,
\een
where $h (x^+)$ is an arbitrary function.  Making the coordinate changes $\gr =
- \log (m z) / m$, $u = x^+$, $v = x^-$, the solution is (\ref{ppw1a}) with $f_1
= 2 h / m^2$, $f_2 = 0$, $f_3 = 0$.


\pa{Gibbons--Pope--Sezgin:}


Gibbons, Pope and Sezgin \ci{gipose} consider supersymmetric solutions of
topologically massive supergravity.  They find that all such solutions are AdS
pp-waves.  We are using their form of the solutions above in (\ref{ppw1a}),
(\ref{ppw1b}) and (\ref{ppw1c}).


\pa{Garbarz--Giribet--V\'{a}squez:}


Garbarz, Giribet and V\'{a}squez \ci{gagiva} obtain solutions for the special
values of $\mu = \pm m$.

The solution in their equation (5), for which $\mu = m$, is
\bea
\ds & = & \fr{r^2}{m^2}  \lt( r^2 - \fr{\gk^2 M}{2 m^2} \rt) ^{-2} 
\ud r^2 - m^2 \lt( r^2 - \fr{\gk^2 M}{2 m^2} \rt) 
\lt( \ud t^2 - \fr{\ud \gf^2}{m^2} \rt) \nnr
&& + \lt[ k \log \lt( \fr{r^2}{r_0^2} - \fr{\gk^2 M}{2 m^2 r_0^2} \rt) 
+ \fr{\gk^2 M}{2} \rt] \lt( \ud t - \fr{\ud \gf}{m} \rt) ^2 \, ,
\eea
where $M$, $r_0$, $\gk$ and $k$ are constants.  Making the coordinate changes
$\ue^{2 m \gr} = m^2 (r^2 - \gk^2 M / 2 m^2)$, $u = t - \gf / m$, $v = - (t +
\gf / m) / 2$, the solution is (\ref{ppw1b}) with $f_1 = 2 k m$, $f_2 = 0$, $f_3
= \gk^2 M / 2 - 2 k \log (m r_0)$.

The solution in their equation (25), for which $\mu = - m$, is
\bea
\ds & = & \fr{r^2}{m^2} \lt( r^2 - \fr{\gk^2 M}{2 m^2} \rt) ^{-2} 
\ud r^2 - m^2 \lt( r^2 - \fr{\gk^2 M}{2 m^2} \rt) \lt( \ud t^2 - 
\fr{\ud \gf^2}{m^2} \rt) \nnr
&& + \lt[ k \lt( r^2 - \fr{\gk^2 M}{2 m^2} \rt) 
\log \lt( \fr{r^2}{r_0^2} - \fr{\gk^2 M}{2 m^2 r_0^2} \rt) + 
\fr{\gk^2 M}{2} \rt] \lt( \ud t + \fr{\ud \gf}{m} \rt) ^2 \, ,
\eea
again with $M$, $r_0$, $\gk$ and $k$ constants.  The same coordinate changes
give (\ref{ppw1c}) with $f_1 = 2 k / m$, $f_2 = - 2 k \log (m r_0) / m^2$, $f_3
= \gk^2 M / 2$.


\ssse{Without cosmological constant}


To take the $m \ra 0$ limit of the general $\mu \neq \pm m$ case
(\ref{ppw1a}), we rewrite the solution in the form
\ben
\ds = \ud \gr^2 + 2 \ue^{2 m \gep \gr} \, \ud u \, \ud v + 
\lt( \ue^{(m \gep - \mu) \gr} f_1 (u) + 
\fr{(\ue^{2 m \gep \gr} - 1)}{2 m \gep} f_2 (u) + f_3 (u) \rt) \ud u^2 \, ,
\een
and then take the limit $\gep \ra 0$, resulting in
\ben
\ds = \ud \gr^2 + 2 \, \ud u \, \ud v + [ \ue^{- \mu \gr} f_1 (u) 
+ \gr f_2 (u) + f_3 (u) ] \, \ud u^2 \, . \la{ppwflat1}
\een
The functions $f_2 (u)$ and $f_3 (u)$ can again be made to vanish by a
coordinate transformation, but are included for comparison with the
literature.


\pa{Martinez--Shepley:}


One of the earliest appearances of any pp-wave solution of TMG appears to be in
an unpublished preprint of Martinez and Shepley \ci{marshe}, which has zero
cosmological constant.  This has been referred to in \ci{nutbak, nutku}.


\pa{Aragone:}


Aragone \cite{aragone} considers a dreibein formalism for TMG.  The solution in
his equation (11) is\footnote{We correct a sign in $g_{uu}$.}
\ben
\ds = \fr{4 N_0^2}{[c(u) - \mu v]^2} \, \ud x^2 + \fr{2}{\mu} \fr{\ud c(u)}{\ud
u} \, \ud u^2 - 2 \, \ud u \, \ud v - 2 N_0 \, \ud u \, \ud x \, ,
\een
where $N_0$ is a constant.  Making the coordinate changes $\ue^{\mu \gr / 2} = v
- c(u)/\mu$, $u' = N_0 (x + v - c(u)/\mu$, $v' = -u + 4 / [\mu^2 v - \mu c(u)]$,
followed by $u' \ra u$ and $v' \ra v$, the solution is (\ref{ppwflat1}) with
$f_1 = 4/\mu^2$, $f_2 = 0$, $f_3 = 0$.


\pa{Percacci--Sodano--Vuorio:}


Percacci, Sodano and Vuorio \cite{Percacci:1986ja} consider stationary solutions
for which the timelike Killing vector has a constant scalar twist.  The solution
in their equation (3.19) is
\ben
\ds = \lt( \fr{2}{\mu x^1} \rt) ^2 (\ud x^1)^2 - \fr{1}{2} \, 
\ud x^0 \, \ud x^2 - \lt( \fr{\mu x^1}{8} \rt) ^2 (\ud x^0)^2 .
\een
In the original expression of the solution, there are functions
$\gw_i$ that are not specified explicitly, but must solve a certain
equation; we have chosen here $\gw_1 = 0$ and $\gw_2 = 16 / \mu^3
(x^2)^2$, but any other choice is equivalent by redefinition of $x^0$.
Making the coordinate changes $u = x^0 / 4$, $\ue^{- \mu \gr / 2} =
x^1$, $v = - x^2$, the solution is (\ref{ppwflat1}) with $f_1 = -
\mu^2 / 4$, $f_2 = 0$, $f_3 = 0$.


\pa{Hall--Morgan--Perj\'{e}s:}


Hall, Morgan and Perj\'{e}s \ci{hamope} look for solutions of TMG with a
particular (Petrov)--Segre type.  The solution in their equation (46) is
\ben
\ds = \ud u^2 + 2 \, \ud x \, \ud r - 2 \ue^{- \mu u} f(x) \, \ud x^2 \, ,
\een
where $f(x)$ is an arbitrary function.  Making the coordinate changes $u \ra
\gr$, $x \ra u$, $r \ra v$, the
solution is (\ref{ppwflat1}) with $f_1 = - 2 f$, $f_2 = 0$, $f_3 = 0$.


\pa{Dereli--Tucker:}


Dereli and Tucker \ci{dertuc} consider solutions with a pp-wave-like ansatz. 
The solution in their equations (2.14) and (2.22) is
\ben
\ds = \ud x^2 + 2 \, \ud u \, \ud v + 2 \lt[ \fr{1}{\mu^2} \ue^{\mu x} 
f_1 (u) + \lt( f_3 (x) - \fr{1}{\mu} f_1 (u) \rt) x + f_2 (u) - 
\fr{1}{\mu^2} f_1 (u) \rt] \ud u^2 \, ,
\een
where $f_1 (u)$, $f_2 (u)$ and $f_3 (u)$ are arbitrary functions.  Making the
coordinate change $\gr = - x$, the solution is (\ref{ppwflat1}), but with the
replacements in (\ref{ppwflat1}) $f_1 \ra 2 f_1 / \mu^2$, $f_2 \ra 2 (f_1 / \mu
- f_3)$, $f_3 \ra 2 (f_2 - f_1 / \mu^2)$.


\pa{Cl\'{e}ment:}


Cl\'{e}ment \cite{clement92} considers stationary rotationally symmetric
solutions.  The solution in his equation (4.15) is
\ben
\ds = \ud r^2 \pm 2 \gs_0 (\ud t - \gw_0 \, \ud \gq) \, \ud \gq - \gs_0 (c
\ue^{\mp \mu r} + b r + a) (\ud t - \gw_0 \, \ud \gq)^2 \, ,
\een
where $a$, $b$, $c$ $\gw_0$ and $\gs_0$ are constants.  Making the coordinate
changes $\gr = \pm r$, $u = t - \gw_0 \gq$, $v = \pm \gw_0 \gq$, the solution is (\ref{ppwflat1}) with $f_1 = - c \gs_0$, $f_2 = 	\mp b \gs_0$, $f_3 = - a
\gs_0$.


\pa{Deser--Steif:}


Deser and Steif \ci{desste} consider an impulsive pp-wave solution.  The
solution in their equation (7) is
\ben
\ds = \ud y^2 - \ud u \, \ud v + \{ 2 \gk^2 E [y + \mu^{-1} (\ue^{- \mu y} - 1)]
\gq(y) \gd(u) + B(u) y + C(u) \} \, \ud u^2 \, ,
\een
where $B(u)$ and $C(u)$ are arbitrary functions, solves $G_{\mu \nu} + \mu^{-1}
C_{\mu \nu} = - \gk^2 T_{\mu \nu}$, with $T_{uu} = E \gd (y) \gd (u)$. 
Considering $y > 0$, replacing $\gd (u) \ra 1$, and then making the coordinate
changes $\gr = y$ and $v \ra - 2 v$, the solution is (\ref{ppwflat1}) with $f_1
= 2 \gk^2 E / \mu$, $f_2 = 2 \gk^2 E + B$, $f_3 = - 2 \gk^2 E / \mu + C$.


\pa{Cavagli\`{a}:}


Cavagli\`{a} \cite{cavaglia} considers certain rotationally symmetric solutions.  The solution in his equations (23) to (26) is
\ben
\ds = - [y(u-v)]^2 \, \ud u \, \ud v - H(u+v) \, \ud \gf \, (\ud u + \ud v) \, ,
\een
where the function $y(u - v)$ is one of: $(\ga / \mu) \tanh [\tf{1}{4} \ga (u - v) - \gb]$, $- (\ga / \mu) \tan[\tf{1}{4} \ga (u - v) - \gb]$, or $(\ga / \mu) /[\tf{1}{4} \ga (u - v) - \gb]$; $\ga$ and $\gb$ are constants.  Making the coordinate changes $\gr = \tf{1}{2} \int \! \ud (u - v) y(u - v)$, $u' = \int \! \ud (u + v) \, H(u + v)$, $v' = - \tf{1}{2} \gf$, and then $u' \ra u$, $v' \ra v$, the solution is of the form (\ref{ppwflat1}).  More explicitly, we have for the respective choices of $y$: $\ue^{\mu \gr / 2} = \cosh [\tf{1}{4} \ga (u - v) - \gb]$, $f_1 = - f_3 = (\ga / 2 H \mu)^2$, $f_2 = 0$; $\ue^{\mu \gr / 2} = \cos [\tf{1}{4} \ga (u - v) - \gb]$, $- f_1 = f_3 = (\ga / 2 H \mu)^2$, $f_2 = 0$; $\ue^{\mu \gr / 2} = \tf{1}{4} \ga (u - v) - \gb$, $f_1 = - (\ga / 2 H \mu)^2$, $f_2 = f_3 = 0$.


\pa{Dereli--Sar\i o\u{g}lu:}


Dereli and Sar\i o\u{g}lu \cite{dersar} consider supersymmetric solutions.  The
solution in their equations (43) to (45) is obtained as a limit of their more
general solution with a non-vanishing cosmological constant discussed
previously.


\pa{Garc\'{\i}a--Hehl--Heinicke--Mac\'{\i}as:}


Garc\'{\i}a, Hehl, Heinicke and Mac\'{\i}as \ci{gahehema} construct a
plane-wave solution to give an example of a type N solution.  The
solution in their equation (124) is
\ben
\ds = \ud y^2 + \ud x^2 - \ud t^2 - 
(B \ue^{\mu y} + A y + C) (\ud t - \ud x)^2 \, ,
\een
where $A$, $B$ and $C$ are constants.  Making the coordinate changes
$\gr = - y$, $u = t - x$, $v = - t - x$, the solution is
(\ref{ppwflat1}) with $f_1 = - B$, $f_2 = A$, $f_3 = - C$.


\pa{Mac\'{\i}as--Camacho:}


Mac\'{\i}as and Camacho \ci{maccam} consider Kerr--Schild solutions of TMG
without a cosmological constant.  They give two solutions explicitly.

The solution in their equation (63) is
\ben
\ds = \ud \xi^2 - 2 \, \ud u \, \ud v + 2 [A \mu^{-1} 
\ue^{\mu (\xi + Y u)} + B (\xi + Y u) + C] (\ud v + Y \, 
\ud \xi + \tf{1}{2} Y^2 \, \ud u)^2 ,
\een
where $Y$ is a constant.  Making the coordinate changes $\gr = - (\xi
+ Y u)$, $u' = v + Y \xi + \tf{1}{2} Y^2 u$, $v' = - u$, and then
$u' \ra u$, $v' \ra v$, the solution is (\ref{ppwflat1}) with $f_1 = 2 A
/ \mu$, $f_2 = - 2 B$, $f_3 = 2 C$.

The solution in their equation (72) is\footnote{They only have two rather than
the three arbitrary functions expected for a third-order theory; this
discrepancy appears to originate in the curvature computation.}
\ben
\ds = \ud \xi^2 - 2 \, \ud u \, \ud v - \mu^{-1} [\ue^{\mu \xi} 
\gc (v) + \ga \xi + C(v) + \ga \mu^{-1}] \, \ud v^2 \, ,
\een
where $\gc (v)$ and $C(v)$ are arbitrary functions, and $\ga$ is a constant. 
Making the coordinate changes $\gr = - \xi$, $u \ra - v$, $v \ra u$, the
solution is (\ref{ppwflat1}) with $f_1 = - \mu^{-1} \gamma (u)$, $f_2 = \mu^{-1}
\ga$, $f_3 = - \mu^{-1} [C(u) + \ga \mu^{-1}]$.


\end{document}